\documentclass[a4paper,11pt]{article}
\usepackage{jheppub}
\usepackage[T1]{fontenc}
\usepackage{lmodern}
\usepackage{microtype}
\usepackage{booktabs}
\usepackage{array}
\usepackage{mathtools}
\usepackage{longtable}
\usepackage{etoolbox}
\usepackage{xcolor}
\newcommand{\la}{\langle}
\newcommand{\ra}{\rangle}

\newcommand{\tr}{\operatorname{tr}}

\allowdisplaybreaks[3]
\makeatletter
\patchcmd{\l@section}{1.0em}{0.3em}{}{}
\makeatother

\renewcommand{\afterTocRuleSpace}{\clearpage}
\title{Cornering the Heterotic String\\[2mm]
\large Higher-Point Consistency of Gauge Theory and Gravity}

\author[a]{Li-Yuan Chiang,}
\emailAdd{li-yuan.chiang@yale.edu}

\author[b,c,d]{Yu-tin Huang,}
\emailAdd{yutinyt@gmail.com}

\author[e]{He-Chen Weng}
\emailAdd{he-chen\_weng@brown.edu}

\definecolor{colorLY}{rgb}{0,0.6,0}

\definecolor{colorHC}{rgb}{1,0,0}

\affiliation[a]{Department of Physics, Yale University, 217 Prospect St, New Haven, CT 06520, USA}
\affiliation[b]{Department of Physics and Center for Theoretical Physics, National Taiwan University, Taipei 10617, Taiwan}
\affiliation[c]{Physics Division, National Center for Theoretical Sciences, Taipei 10617, Taiwan}
\affiliation[d]{Max Planck{-}IAS{-}NTU Center for Particle Physics, Cosmology and Geometry, Taipei 10617, Taiwan}
\affiliation[e]{Department of Physics, Brown University, Providence, RI 02912, USA}

\abstract{We study tree-level completions of four-dimensional $\mathcal N=4$ Yang--Mills theory coupled to $\mathcal N=4$ supergravity. Six-point factorization and supersymmetry Ward identities, supplemented by a peculiar-parity restriction on scalar contact terms, impose nonlinear relations between the single- and double-trace Wilson coefficients, $a_{k,q}$ and $b_{k,q}$, of the four-gluon amplitude. The simplest relation, $\kappa^2a_{0,0}=0$, excludes the single-trace $F^4$ interaction at nonzero gravitational coupling, despite its compatibility with four-point supersymmetry. On this gravitational branch, five double-trace parameters determine all coefficients through $a_{7,q}$ and $b_{8,q}$, which are reproduced by a common exponential generating form. Both the heterotic string and a family of infinite-spin-tower (IST) amplitudes realize this structure. Combining these relations with dispersion relations and partial-wave positivity in $SO(N)$ color channels yields a narrow numerical band closely bounded by the heterotic and simplest IST amplitudes. A spin-dependent mass gap excludes the IST spectrum and further localizes the region near the heterotic trajectory. An analytic unitarity analysis of the simplest IST gives $N\leq48$ and $g^2/(\kappa^2 m^2)\leq4$, matching the upper limits of the heterotic amplitude, where $m$ is the mass of the first massive level. Along the heterotic trajectory, the numerical bounds also sharply constrain the gauge-to-gravity coupling ratio at $N=48$.}
\begin{document}
\maketitle
\flushbottom

\section{Introduction}\label{sec:intro}

Can perturbative string theory be derived from low-energy consistency without assuming a worldsheet description of its ultraviolet completion? This question is a central motivation for the effective field theory (EFT) bootstrap. Analyticity, crossing symmetry, unitarity, and suitable high-energy boundedness constrain the Wilson coefficients of low-energy scattering amplitudes, delineating the boundaries of \textit{theory space}~\cite{deRham:2017avq,Bellazzini:2020cot,Caron-Huot:2020cmc,Tolley:2020gtv,ArkaniHamedHuangHuang2021EFT,Sinha:2020win,Caron-Huot:2021rmr}. Yet four-point positivity bounds generally allow a much larger region than that occupied by known ultraviolet completions. The small islands populated by explicit field-theory and string-theory examples suggest that further consistency conditions remain to be identified~\cite{BernKosmopoulosZhiboedov2021,Chiang:2022jep}.

One approach supplements the bootstrap with relations exhibited by string amplitudes. Combining open-string monodromy relations with positivity produces small regions that converge numerically towards the Veneziano amplitude~\cite{Huang:2020nqy,Chiang:2023quf,Berman:2023jys}, an identification subsequently established analytically in~\cite{Wan:2026pjq}. Hidden zeros and higher-point splitting identities provide another source of nonlinear constraints: combined with unitarity and appropriate spectral assumptions, they isolate shrinking islands around string amplitudes~\cite{Berman:2025owb}. Complementary analytic constructions establish uniqueness from assumptions about residue zeros and high-energy softness~\cite{Cheung:2025tbr}. These results, however, rely on substantial ``stringy'' input that lacks a purely infrared motivation.

An important step towards an infrared derivation was taken in~\cite{Elvang:2026pmc}, which showed that six-point supersymmetry Ward identities, supplemented by a scalar-parity condition, impose nonlinear relations among the four-point Wilson coefficients of maximally supersymmetric Yang--Mills (SYM) theory. Through the orders examined, these relations imply the monodromy constraints and, together with unitarity, determine the coefficients to high precision in agreement with the Veneziano amplitude. The scalar-parity condition, subsequently termed \emph{peculiar parity}, is a reflection property of selected tree-level scalar amplitudes. Assuming $CP$ invariance, it is equivalent in that sector to separate invariance under spacetime parity and conjugation of the R-symmetry quantum numbers. The analogous analysis of maximally supersymmetric gravity yields a nonconvex allowed region with the Virasoro--Shapiro amplitude and an infinite-spin-tower amplitude at its corners~\cite{Berman:2026ezk}.

Although these analyses examine only finitely many Wilson coefficients, recent work has established the identification to all orders. Special six-point collinear limits of the supersymmetry Ward identities, together with scalar parity, eliminate the contact-term ambiguity and yield a quadratic functional identity for the four-point function. Combining this identity with a moment interpretation of the dispersive sum rules establishes the uniqueness of the Veneziano and Virasoro--Shapiro amplitudes, with an additional finite-spin condition at the lowest massive pole in the gravitational case~\cite{Sung:2026euf,Huang:2026ghg}.

We extend this program to gauge theory coupled to gravity, studying tree-level completions of four-dimensional $\mathcal N=4$ Einstein--Yang--Mills (EYM) theory with gauge group $SO(N)$. The heterotic string completes both interactions at the same perturbative order, but a recent four-point bootstrap study of ten-dimensional half-maximal theories did not isolate its amplitudes~\cite{BoisvertKnopRastelli}. Here we include the constraints from parity-restricted supersymmetric six-point amplitudes.

The four-gluon amplitude contains single- and double-trace reduced functions, with contact coefficients $a_{k,q}$ and $b_{k,q}$ and leading massless poles $-g^2/(st)$ and $-\kappa^2/s$, respectively. Here $g$ and $\kappa$ are the gauge and gravitational couplings. Products of the four-point amplitudes determine the three-particle factorization residues at six points. Color decomposition separates these products into three sectors: the single-trace sector involves two single-trace four-point amplitudes, the $\tr(T^2)\tr(T^4)$ sector mixes single- and double-trace amplitudes, and the $\tr(T^2)^3$ sector involves two double-trace amplitudes.\footnote{We do not include the $\tr(T^3)\tr(T^3)$ sector, which requires additional amplitudes with external gravitational states.} We construct scalar components with the required residues and impose the NMHV Ward identities on components related by trace-preserving permutations. Restricting the local six-scalar contact terms to parity-even Mandelstam polynomials in the pole decomposition specified below, we eliminate their coefficients and the remaining five-point freedom to obtain nonlinear compatibility conditions on the four-point data.

The simplest mixed-sector condition is
\begin{equation}\label{eq:intro-F4-obstruction}
    \boxed{\kappa^2 a_{0,0}=0\,.}
\end{equation}
It separates two nondegenerate branches: a gravity-decoupled branch with $a_{0,0}\neq0$, on which the double-trace coefficients vanish through the orders studied and the constraints reduce to the open-superstring system, and a gravitational branch with $\kappa\neq0$ and $a_{0,0}=0$. We refer to these as the type-I and heterotic branches. On the latter, the nonlinear constraints also require the following single-trace coefficients to vanish:
\begin{equation}
a_{2,0}=a_{2,1}=a_{4,0}=a_{6,0}=0\,.
\end{equation}
These zeros do not follow from supersymmetry alone, since the corresponding coefficients are nonzero on the open-string branch; they require both supersymmetry and scalar parity.

Combining the single-trace relations of ref.~\cite{Elvang:2026pmc} with our mixed- and triple-trace relations, we solve the system through $a_{7,q}$ and $b_{8,q}$. On the heterotic branch, five normalized ``diagonal'' double-trace coefficients determine all retained data,
\begin{equation}\label{eq:intro-seeds}
    \left\{\widehat b_{0,0},\widehat b_{2,2},\widehat b_{4,4},
    \widehat b_{6,6},\widehat b_{8,8}\right\},\qquad
    \widehat a_{k,q}=\frac{a_{k,q}}{g^2},\quad
    \widehat b_{k,q}=\frac{b_{k,q}}{\kappa^2}.
\end{equation}
The solution through these orders is generated by the compact form
\begin{equation}\label{eq:intro-generating-form}
\begin{aligned}
    F^{\mathrm{ST}}(s,t)&=-\frac{g^2}{st}\,e^{E(s,t,u)},&
    F^{\mathrm{DT}}(s,t)&=-\frac{\kappa^2}{s(1+\widehat b_{0,0}s)}\,e^{E(s,t,u)},\\
    E(s,t,u)&=\sum_{r\geq1}\frac{4^r\widehat b_{2r,2r}}{2r+1}
    \bigl(s^{2r+1}+t^{2r+1}+u^{2r+1}\bigr),&s+t+u&=0\,,
\end{aligned}
\end{equation}
where $F^{\mathrm{ST}}$ and $F^{\mathrm{DT}}$ are the single- and double-trace four-point functions. The heterotic string realizes this structure with $\widehat b_{0,0}=\alpha'$ and $\widehat b_{2r,2r}=2^{1-2r}\zeta(2r+1)\alpha'^{2r+1}$; unitarity must constrain the remaining independent coefficients.

A family of infinite-spin-tower (IST) amplitudes also satisfies the retained algebraic conditions, analogous to the solutions appearing in the gravitational bootstrap~\cite{Berman:2026ezk}. Their coefficients follow by replacing odd zeta values with generalized harmonic numbers. For a positive integer $q$, the amplitude has $q$ massive levels with infinite spin support, while the limit $q\to\infty$ reproduces the heterotic amplitude. The simplest member, $q=1$, has a single massive level of mass $m$. Its partial-wave expansion in color projectors gives the analytic unitarity bounds $N\leq48$ and $g^2/(\kappa^2m^2)\leq4$ in four dimensions, together with an $N$-dependent lower bound on the coupling ratio. Both upper bounds coincide with those of the heterotic four-gluon amplitude.

We constrain the remaining parameters using dispersion relations and positivity in irreducible color channels. With several trace structures present, positivity applies to the exchanged representations, not independently to each trace coefficient. Color crossing and gravitational exchange have also been used to derive spectral constraints and scattering-based completeness results~\cite{HillmanHuangRodinaRumbutis2024,CalistoEtAl2025Completeness}. Our numerical implementation uses a hierarchy of relaxations: fixing selected low-energy parameters makes the corresponding relations linear, allowing each point in an external-parameter scan to be tested by semidefinite programming. Higher-dimensional scans incorporate more of the nonlinear constraints.

\begin{figure}[htpb]
    \centering
    \includegraphics[width=0.85\linewidth]{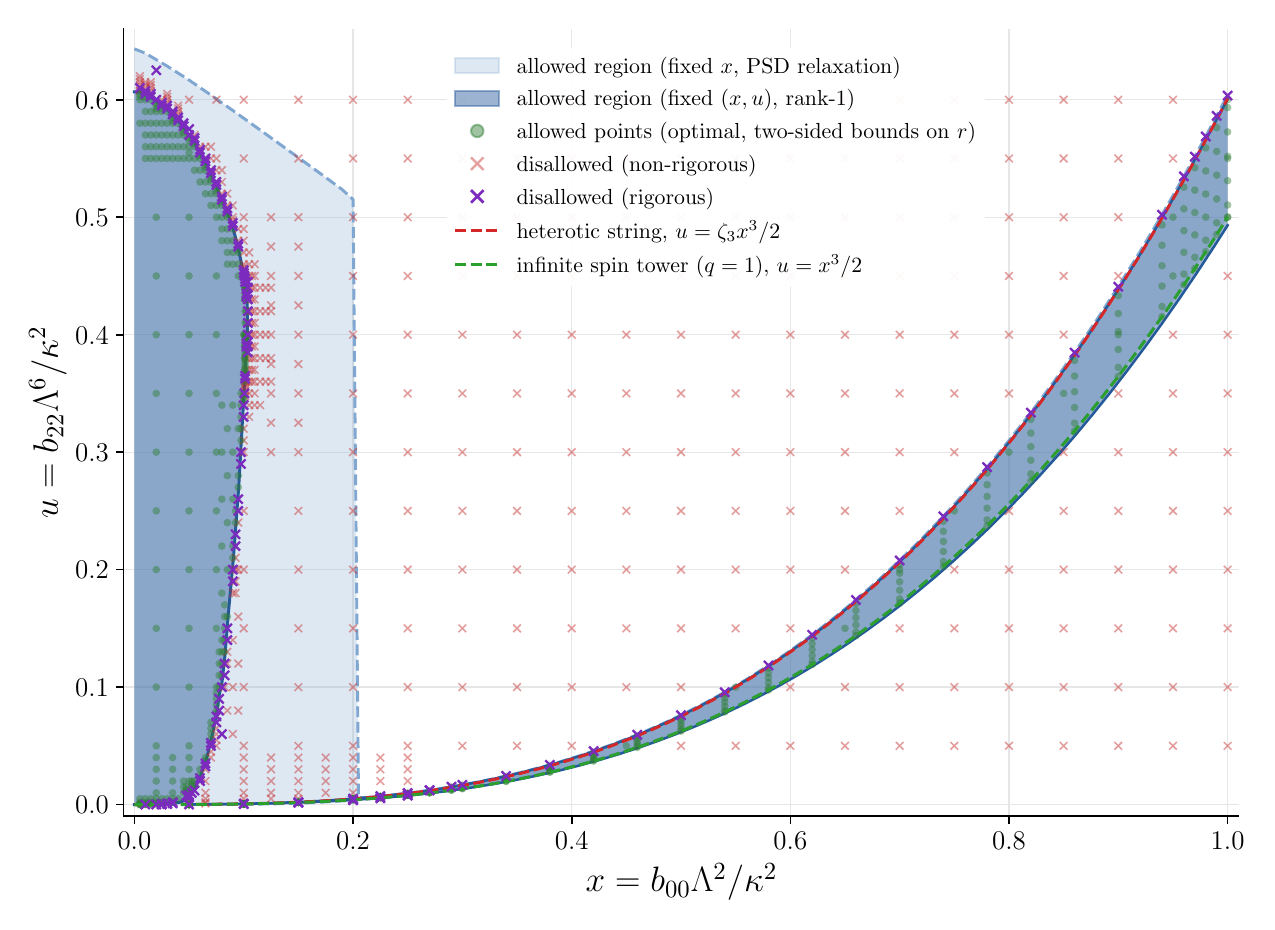}
    \caption{
		Allowed region in $(b_{00},b_{22})$ at $N=10$, normalized against the EFT cutoff $\Lambda$ and the gravitational coupling $\kappa$, focusing on the region $x \geq 0$. Rescalings of the heterotic and IST solutions generate cubic trajectories that closely approach the boundaries of the band on the right.
    }
    \label{fig:Intro}
\end{figure}

The allowed region in the $(\widehat b_{0,0},\widehat b_{2,2})$ plane contains a narrow band whose upper and lower boundaries are closely approached by rescalings of the heterotic and $q=1$ IST amplitudes,
\begin{equation}\label{eq:intro-cubic-trajectories}
    \widehat b_{2,2}=\frac{\zeta(3)}{2}\widehat b_{0,0}^{\,3}
    \quad\hbox{and}\quad
    \widehat b_{2,2}=\frac{1}{2}\widehat b_{0,0}^{\,3},
\end{equation}
respectively, as shown in Figure~\ref{fig:Intro}. A broader region near the vertical axis shrinks as $N$ increases. At $N=49$, the extended band disappears from the numerical bounds, leaving only a small region near the origin. A spin-dependent mass gap excludes the IST spectrum and further narrows the band; at the slope saturated by the heterotic spectrum, it lies extremely close to the heterotic trajectory. Along this trajectory, the bootstrap also constrains the gauge-to-gravity coupling ratio. At $N=48$, the mass-rescaling-invariant combination $\widehat b_{0,0}g^2/\kappa^2$ lies in a narrow interval containing the heterotic value $4$.

Section~\ref{sec:color-bootstrap} introduces the color-dressed $\mathcal N=4$ superamplitudes and lower-point EFT data. Section~\ref{sec:ward-derivation} constructs the six-point scalar amplitudes and implements the Ward identities and parity restriction. Section~\ref{sec:constraints} presents the nonlinear constraints, their two nondegenerate branches, and the gravitational solution. Sections~\ref{sec:branches} and~\ref{sec:results} develop the dispersive bootstrap and present the numerical bounds and IST amplitudes. Appendix~\ref{sec:single_trace_limit} relates single-trace positivity to color-projector positivity, and appendix~\ref{sec:Unitarity_IST} gives the analytic unitarity analysis of the $q=1$ IST amplitude. The nonlinear SUSY constraints are supplied in an ancillary file.

\section{Color-dressed superamplitudes for $\mathcal N=4$ EYM EFTs}\label{sec:color-bootstrap}

We review the massless on-shell superspace and three- and four-point superamplitudes of $\mathcal N=4$ EYM, then construct the five-gluon superamplitude by matching its massless factorization residues.

\subsection{On-shell superspace for $\mathcal N=4$ EYM EFTs}
\label{sec:color-superamps}

In four-dimensional chiral superspace, the on-shell SUSY generators are
\begin{equation}
    \tilde{Q}_a=\sum_i|i\rangle \eta_{i,a},\quad Q^a=\sum_i[i| \partial_i^a,\quad 1\leq a \leq 4.
\end{equation}
Here $\eta_{ia}$ are Grassmann-odd variables, with $i$ labeling the external leg and $a$ the $SU(4)_R$ index, and $\partial_i^a=\partial/\partial\eta_{ia}$. The SUSY Ward identities require
\begin{equation}
    \tilde{Q}_a \mathcal{A}_n = Q^a \mathcal{A}_n =0.
\end{equation}
The condition $\tilde{Q}_a\mathcal A_n=0$ is solved by factoring out a Grassmann delta function,
\begin{equation}
\begin{aligned}
    &\mathcal{A}_n^{\text{N}^k\text{MHV}} = \delta^{(8)}(\tilde{Q})P_n^{4k},\quad \delta^{(8)}(\tilde{Q}) = \prod_{a=1}^4\sum_{i<j}\langle ij\rangle\eta_{ia}\eta_{ja},
\end{aligned}
\end{equation}
where $P_n^{4k}$ is a degree-$4k$ polynomial in the Grassmann variables.

The gluon multiplet and the positive- and negative-helicity graviton multiplets are
\begin{equation}
\begin{aligned}
    G_i &= g_i^+ + \eta_{ia} \lambda_i^a + \eta_{ia}\eta_{ib} z_i^{ab} + \eta_{ia}\eta_{ib}\eta_{ic}\lambda_i^{abc} + \eta_{ia}\eta_{ib}\eta_{ic}\eta_{id}g_i^-,\\
    H_i^+ &= h_i^+ + \eta_{ia} \psi_i^{+a} + \eta_{ia}\eta_{ib} v_i^{+ab} + \eta_{ia}\eta_{ib}\eta_{ic}\chi_i^{+abc} + \eta_{ia}\eta_{ib}\eta_{ic}\eta_{id}\phi_i,\\
    H_i^- &= \bar{\phi}_i + \eta_{ia} \chi_i^{-a} + \eta_{ia}\eta_{ib} v_i^{-ab} + \eta_{ia}\eta_{ib}\eta_{ic}\psi_i^{-abc} + \eta_{ia}\eta_{ib}\eta_{ic}\eta_{id}h_i^-.
\end{aligned}
\end{equation}
The gluon multiplet contains the gluons $g_i^\pm$, gluinos $\lambda_i^a$ and $\lambda_i^{abc}$, and scalars $z_i^{ab}$ in the $\mathbf 6$ of $SU(4)_R$. The $\mathcal N=4$ supergravity states split between $H_i^+$, containing the positive-helicity graviton, gravitinos, and graviphotons, and its CPT conjugate $H_i^-$. In flat space, the graviphotons are abelian and do not mix with the nonabelian SYM gluons. 

Component amplitudes are extracted from the superamplitude using the Grassmann derivatives $\partial_i^a$. The $SU(4)_R$ symmetry restricts the Grassmann degree to multiples of four, giving the decomposition
\begin{equation}
    \mathcal{A}_n = \mathcal{A}_n^{\text{MHV}} + \mathcal{A}_n^{\text{NMHV}}+\mathcal{A}_n^{\text{N}^2\text{MHV}} + \dots,
\end{equation}
where $\mathcal A_n^{\text{N}^k\text{MHV}}$ has Grassmann degree $4k+8$, with $0\leq k\leq n-4$ for $n\geq 4$. For $n=3$, the amplitude has two sectors, $\mathcal{A}_3 = \mathcal{A}_3^{\overline{\text{MHV}}} + \mathcal{A}_3^{\text{MHV}}$, where the former has Grassmann degree 4 and the latter has Grassmann degree 8. For a pure-gluon component amplitude, the number of negative-helicity gluons is equal to the Grassmann degree divided by four.

\subsection{Three- and four-point amplitudes}

For gluon scattering, we use the trace basis of $SO(N)$; the construction can be adapted to other gauge groups. Supersymmetry fixes the three-point amplitudes and constrains the four-gluon amplitude up to polynomial terms. The three-gluon and two-gluon--one-graviton superamplitudes are
\begin{equation}
\begin{aligned}
    \mathcal{A}[G^3]^{\text{MHV}}&=\sqrt{2}g\frac{\delta^{(8)}(\tilde{Q})}{\la12\ra\la23\ra\la31\ra} \tr(123), &\mathcal{A}[G^3]^{\overline{\text{MHV}}}= \sqrt{2}g\frac{\delta^{(4)}(m_{123,a})}{[12][23][31]} \tr(123),\\
     \mathcal{A}[H^+GG]^{\overline{\text{MHV}}}&= \kappa \frac{\delta^{(4)}(m_{123,a})}{[23]^2} \tr(23), &\mathcal{A}[H^-GG]^{\text{MHV}}= \kappa \frac{\delta^{(8)}(\tilde{Q})}{\la23\ra^2} \tr(23),
\end{aligned}
\end{equation}
where $\tr(i,\dots,j) = \tr(T^{a_i}\dots T^{a_j})$.
We use the convention $\tr(ij)=\delta^{ij}$ and the shorthand
\begin{equation}\label{eq:grassmann_mijka}
    \delta^{(4)}(m_{123,a}) = \prod_{a=1}^4m_{123,a},\qquad m_{ijk,a} = [ij]\eta_{ka}+[jk]\eta_{ia}+[ki]\eta_{ja}.
\end{equation}
The normalization of $\tr(ij)=\delta^{ij}$ leads to the completeness relation
\begin{equation}
    \sum_{a}(T^a)_{ij}(T^a)_{kl} = \frac{1}{2}\left(\delta_{il}\delta_{jk}-\delta_{ik}\delta_{jl}\right).
\end{equation}

At four points, the trace basis contains single- and double-trace structures. Supersymmetry fixes the common kinematic prefactor $\delta^{(8)}(\tilde Q)[34]^2/\la12\ra^2$, where the spinor brackets account for the helicity $+1$ of the gluon superfield.\footnote{Other spinor-bracket choices are equivalent: $\frac{[34]^2}{\la12\ra^2}=\frac{[12]^2}{\la34\ra^2}=\frac{[13]^2}{\la24\ra^2}$.} The dynamics is encoded in the reduced scalar function $A_4$, defined by $\mathcal A[G^4]=\delta^{(8)}(\tilde Q)[34]^2/\la12\ra^2\,A_4$, with
\begin{equation}\label{eq:single_double_tr}
\begin{aligned}
A_4
&=\tr(1234) F^{\mathrm{ST}}(s,t)
 +\tr(1342) F^{\mathrm{ST}}(u,s)
 +\tr(1423) F^{\mathrm{ST}}(t,u)\\
&+\tr(12)\tr(34) F^{\mathrm{DT}}(s,t)
 +\tr(13)\tr(24) F^{\mathrm{DT}}(u,s)
 +\tr(14)\tr(23) F^{\mathrm{DT}}(t,u).
\end{aligned}
\end{equation}
Here $s=(p_1+p_2)^2$, $t=(p_1+p_4)^2$, and $u=(p_1+p_3)^2$, with all momenta outgoing. The low-energy expansions are
\begin{align}
F^{\mathrm{ST}}(s,t)&=-\frac{g^2}{st} +f^{\mathrm{ST}}(s,t),&
f^{\mathrm{ST}}(s,t)&=\sum_{k\ge0}\sum_{q=0}^k a_{k,q}s^{k-q}t^q,
\label{eq:ST-convention}\\
F^{\mathrm{DT}}(s,t)&=-\frac{\kappa^2}{s} +f^{\mathrm{DT}}(s,t),&
f^{\mathrm{DT}}(s,t)&=\sum_{k\ge0}\sum_{\substack{0\le q\le k\\q\ {\rm even}}}
 b_{k,q}s^{k-q}(t-u)^q\,.
\label{eq:DT-convention}
\end{align}
The massless poles of $F^{\mathrm{ST}}$ occur in both arguments and describe gluon exchange, whereas those of $F^{\mathrm{DT}}$ occur only in its first argument and describe graviton exchange. Cyclicity gives $a_{k,q}=a_{k,k-q}$, while the double-trace exchange symmetry $F^{\mathrm{DT}}(s,t)=F^{\mathrm{DT}}(s,u)$ requires $b_{k,q}=0$ for odd $q$.

\noindent \textbf{Examples}

\begin{itemize}
\item Type-I superstring: The disk amplitude contains only single-trace terms, with reduced function
\begin{equation}
F_{\mathrm{open}}^{\mathrm{ST}}(s,t)
=-g^2\alpha'^2
\frac{\Gamma(-\alpha's)\Gamma(-\alpha't)}
{\Gamma(1-\alpha's-\alpha't)}.
\end{equation}
This yields
\begin{equation}
\begin{aligned}
a_{0,0}&=g^2\alpha'^2\zeta_2,&a_{1,0}&=g^2\alpha'^3\zeta_3,\\
a_{2,0}&=g^2\alpha'^4\zeta_4,&a_{2,1}&=\tfrac14g^2\alpha'^4\zeta_4,\\
a_{3,0}&=g^2\alpha'^5\zeta_5,&
a_{3,1}&=g^2\alpha'^5(2\zeta_5-\zeta_2\zeta_3).
\end{aligned}
\end{equation}

\item Heterotic string: 
The two reduced functions of the heterotic amplitude can be expressed in terms of the Virasoro--Shapiro form factor
\begin{equation}\label{eq:heterotic_factor}
V(s,t,u)=-\frac{\Gamma(-\alpha' s)\Gamma(-\alpha' t)\Gamma(-\alpha' u)}
{\Gamma(\alpha' s)\Gamma(\alpha' t)\Gamma(\alpha' u)},
\end{equation}
as
\begin{equation}\label{eq:heterotic_trace_functions}
F_{\mathrm{Het}}^{\mathrm{ST}}(s,t)
=-\frac{g^2}{st} V(s,t,u),\qquad
F_{\mathrm{Het}}^{\mathrm{DT}}(s,t)
=-\frac{\kappa^2}{s(1+\alpha's)} V(s,t,u).
\end{equation}
The other color orderings follow from \eqref{eq:single_double_tr}.
Its low-energy expansion is generated by~\cite{Schlotterer:2012ny}
\begin{equation}\label{eq:dt_generating}
V=e^{\mathcal E},\qquad
\mathcal E=\sum_{r=1}^{\infty}\frac{2\zeta_{2r+1}\alpha'^{2r+1}}{2r+1}
\left(s^{2r+1}+t^{2r+1}+u^{2r+1}\right).
\end{equation}
Through the first nontrivial orders this gives
\begin{equation}\label{eq:heterotic_coefficients}
\begin{aligned}
&a_{0,0}=0,\quad &a_{1,0}&=2g^2\alpha'^3\zeta_3,\quad &a_{2,0}&=0,\\
&a_{2,1}=0,\quad &a_{3,0}&=2g^2\alpha'^5\zeta_5,\quad &a_{3,1}&=4g^2\alpha'^5\zeta_5,\\
&a_{4,0}=0,\quad &a_{4,1}&=-2g^2\alpha'^6\zeta_3^2,\quad & a_{4,2}&=-4g^2\alpha'^6\zeta_3^2,\\
&b_{0,0}=\kappa^2\alpha',\quad &b_{1,0}&=-\kappa^2\alpha'^2,\quad &b_{2,0}&=\kappa^2\alpha'^3(1-\tfrac12\zeta_3),\\
&b_{2,2}=\tfrac12\kappa^2\alpha'^3\zeta_3,\quad &b_{3,0}&=-\kappa^2\alpha'^4(1-\tfrac12\zeta_3),\quad &b_{3,2}&=-\tfrac12\kappa^2\alpha'^4\zeta_3.
\end{aligned}
\end{equation}
Several heterotic coefficients vanish, including $a_{0,0}$, although their type-I counterparts are nonzero and both amplitudes obey four-point supersymmetry. Through the orders checked below, these zeros follow from higher-point SUSY Ward identities and scalar parity in the presence of tree-level gravitational coupling.

\end{itemize}

\subsection{Five-point amplitudes}

At five points, only the MHV and $\overline{\mathrm{MHV}}$ sectors occur, so it suffices to consider the MHV superamplitude,
\begin{equation}
    \mathcal{A}[G^5]^{\text{MHV}}=\sum_{\sigma\in S_4/ Z_2}\frac{1}{\sqrt{2}}\tr(1\sigma(2345))\mathcal{A}[1\sigma(2345)]+\sum_{\substack{i<j,\,k<l<m}}\sqrt{2}\tr(ij)\tr(klm)\mathcal{A}[ij;klm],
\end{equation}\color{black}
where $\{i,j,k,l,m\}$ are distinct integers from 1 to 5. The single-trace residues do not involve the four-gluon double-trace amplitude and therefore do not generate nonlinear constraints involving $b_{k,q}$. We refer to~\cite{Elvang:2026pmc} for the single-trace analysis.

The double-trace partial amplitude $A[12;345]$ can have poles only at $\{\la12\ra,\la34\ra,\la45\ra,\la53\ra\}$. Using a common denominator, we write
\begin{equation}
\begin{aligned}
    &\mathcal{A}[12;345] = \delta^{(8)}(\tilde{Q})\frac{N_5[12;345]}{\la12\ra\la34\ra\la45\ra\la53\ra}\,.
\end{aligned}
\end{equation}
Little-group covariance requires $N_5$ to carry little group weight $1/2$ on legs 1 and 2. We use the basis
\begin{equation}
\begin{aligned}\label{eq:N5_ansatz}
    &N_5[12;345] = [12] N_5^{(1)}[12;345] + [13]\la34\ra[42] N_5^{(2)}[12;345]\,,
\end{aligned}
\end{equation}
where $N_5^{(1)}$ and $N_5^{(2)}$ are polynomials in the independent five-point Mandelstam variables $\{s_{12},s_{23},s_{34},s_{45},s_{15}\}$. Invariance of $A[12;345]$ under $1\leftrightarrow2$ and cyclic permutations of $(345)$ requires the numerator to be antisymmetric under the former and invariant under the latter.

At $\la34\ra=0$, $\la45\ra=0$, or $\la53\ra=0$, the amplitude factorizes into a three-gluon $\overline{\mathrm{MHV}}$ amplitude and a four-gluon double-trace amplitude. Cyclicity of $(345)$ reduces the matching to one pole,
\begin{equation}
    \mathcal{A}[12;345]\bigg|_{\la45\ra\to 0} = \delta^{(8)}(\tilde{Q})\left(\frac{-\kappa^2 g [12]}{\la12\ra\la34\ra\la45\ra\la53\ra}+\frac{-g [12]^2}{\la34\ra\la45\ra\la53\ra}f^{\text{DT}}(s_{12},s_{23})\right).
\end{equation}
The leading terms of the numerator are fixed by the lower-point coefficients $N_5[12;345]=-\kappa^2g[12]+g\,b_{0,0}s_{12}[12]+O(p^5)$ while the first free parameters that are not determined by residue matching first appear at order $p^7$. At $\la12\ra=0$, the amplitude instead factorizes into a gluon--gluon--graviton amplitude and a three-gluon--one-graviton amplitude. Since the latter is outside our scope, we leave this residue unconstrained.

Table~\ref{tab:N5_residue_matching} lists the remaining numerator parameters at each order after residue matching. We solve the matching conditions at random rational kinematics using the null-space solver~\cite{SparseRREF}.
\begin{table}[h!]
    \centering
    \begin{tabular}{|c|c|c|c|c|c|c|c|c|c|c|c|}
        \hline
        Order in $N_5$ & $p^1$ & $p^3$ & $p^5$ & $p^7$ & $p^9$ & $p^{11}$ & $p^{13}$ & $p^{15}$ & $p^{17}$ & $p^{19}$ & $p^{21}$ \\
        \hline
        Ansatz & 1 & 1 & 3 & 9 & 18 & 32 & 58 & 90 & 137 & 203 & 287 \\
        \hline
        Matching residue $\la45\ra \to0$ & 0 & 0 & 0 & 1 & 3 & 7 & 19 & 34 & 60 & 98 & 150\\
        \hline
    \end{tabular}
    \caption{Number of free parameters in the five-point numerator at each order. The ansatz obeys cyclicity in $(345)$ and antisymmetry under $1\leftrightarrow2$. The final row counts parameters left undetermined by the four-point data after residue matching.}\label{tab:N5_residue_matching}
\end{table}

\section{Six-point EFT amplitudes and SUSY Ward identities}\label{sec:ward-derivation}

At six points, the MHV and N$^2$MHV sectors are CPT conjugates, and the MHV Ward identities are solved by $\delta^{(8)}(\tilde Q)$. We therefore seek constraints on the four-point Wilson coefficients in the NMHV sector, whose color decomposition contains four trace structures:
\begin{equation}
\begin{aligned}
\mathcal{A}[G^6]^{\text{NMHV}}&=\sum_{\sigma\in S_5/ Z_2}\frac{1}{2}\tr(1\sigma(23456))\mathcal{A}[1\sigma(23456)]\\
&+\sum_{\substack{i<j,\,k<l<m<n\\\sigma\in S_3/ Z_2}}
 \tr(ij)\tr(k\sigma(lmn))\mathcal{A}[ij;k\sigma(lmn)]\\
&+\sum_{\substack{j<k,\,l<m<n}}
 2\tr(1jk)\tr(lmn)
 \mathcal{A}[1jk;lmn]\\
&+\sum_{\substack{k<l,\,m<n\\k<m}}
 \tr(1j)\tr(kl)\tr(mn)\mathcal{A}[1j;kl;mn],
\end{aligned}
\end{equation}
In the second sum, $\{i,j,k,l,m,n\}$ are distinct integers from 1 to 6; in the third and fourth sums, $\{j,k,l,m,n\}$ are distinct integers from 2 to 6.

\subsection{SUSY Ward identities}

The six-point NMHV superamplitude takes the form
\begin{equation}
\mathcal{A}_6^{\rm NMHV}=\delta^{(8)}(\tilde{Q})\mathcal{P}_6
\end{equation}
where $\mathcal P_6$ has Grassmann degree four and obeys $Q\mathcal P_6=0$. For each R-symmetry index, $\delta^{(2)}(\tilde Q)$ and the $Q$ constraint eliminate the Grassmann variables on four legs. Distributing the four remaining $\eta$ variables between the other two legs gives five independent structures. The superamplitude can therefore be written in terms of five component amplitudes~\cite{Elvang:2009wd,Elvang:2010xn},
\begin{equation}
    \mathcal{A}_6^{\text{NMHV}} = A_1 \mathcal X_{1111} + A_2 \mathcal X_{(1112)} + A_3 \mathcal X_{(1122)} + A_4 \mathcal X_{(1222)} + A_5 \mathcal X_{2222},
\end{equation}
where $A_i$ are component amplitudes and
\begin{equation}
\begin{aligned}
    \mathcal X_{ijkl} &= \delta^{(8)}(\tilde{Q})\frac{m_{i34,1}m_{j34,2}m_{k34,3}m_{l34,4}}{[34]^4\la56\ra^4},\\
\end{aligned}
\end{equation}
where $m_{ijk,a}$ is given in \eqref{eq:grassmann_mijka}. Parentheses in $\mathcal X_{(ijkl)}$ denote a sum over inequivalent permutations. We use the package~\cite{Headrick:grassmann} for Grassmann-variable manipulations in this work. 

A convenient implementation projects the superamplitude onto scalar components~\cite{Elvang:2026pmc}. In the single-trace construction, one starts with $Z_1=A[1^{12}2^{13}3^{14}4^{23}5^{34}6^{24}]$ and obtains $Z_r$, $r=1,\ldots,6$, by cyclically shifting the scalar assignment by $r-1$ positions while keeping the ordered momentum slots fixed. For example, $Z_2=A[1^{13}2^{14}3^{23}4^{34}5^{24}6^{12}]$. The projections give $Z_r=\sum_{i=1}^5C_{ri}A_i$, with $C_{ri}$ determined by the $\mathcal X$ polynomials. At generic kinematics, the first five projections determine the $A_i$, and the sixth yields a Ward identity $Z_6=\sum_{r=1}^5c_rZ_r$ with purely kinematic coefficients. Only $Z_1$ needs to be constructed explicitly, since cyclic momentum relabeling generates the other five components.

Since supersymmetry acts trivially on color, we apply this construction independently in each trace sector, choosing scalar components related by trace-preserving permutations. We match three-particle residues to products of four-point amplitudes and, where present, two-particle residues to products of three- and five-point amplitudes, then add local contact terms. Imposing the scalar Ward identities order by order, with the parity restriction specified below, constrains the contact terms and yields nonlinear relations among the four-point coefficients.

The single-trace sector, which constrains $a_{k,q}$, was analyzed in~\cite{Elvang:2026pmc}. The $\tr(T^3)\tr(T^3)$ sector requires amplitudes with three gluons and one graviton, introducing EFT data beyond our scope. We focus on $\tr(T^2)\tr(T^4)$, whose three-particle residues mix single- and double-trace amplitudes, and $\tr(T^2)^3$, whose residues involve two double-trace amplitudes. These sectors constrain $a_{k,q}$ and $b_{k,q}$ jointly, and $b_{k,q}$ alone, respectively.

\subsection{Six-point amplitude ans\"atze}

We construct scalar-component ans\"atze in the $\tr(T^2)\tr(T^4)$ and $\tr(T^2)^3$ sectors by matching their massless factorization residues.

\subsubsection{The $\tr(T^2)\tr(T^4)$ sector}

We choose $A[1^{12}2^{13};3^{14}4^{23}5^{34}6^{24}]$ and the seven other scalar assignments generated by $(3456)$ and $(12)$. This component has poles at $s_{345}=0$, $s_{346}=0$, and $s_{34}=0$. Matching the three-particle residues and the double poles $s_{34}=s_{345}=0$ and $s_{34}=s_{346}=0$ gives 
\begin{equation}
\begin{aligned}
    &A[1^{12} 2^{13};3^{14}4^{23} 5^{34} 6^{24}] = \kappa^2 g^2 \left(-\frac{s_{235}}{s_{34}}+\frac{s_{25}s_{46}}{s_{34}s_{346}}+\frac{s_{16}s_{35}}{s_{34}s_{345}}\right)\\
    &- \kappa^2\frac{s_{25}s_{36}s_{46}}{s_{346}}f^{\text{ST}}(s_{34},s_{36}) 
    -g^2\frac{s_{25}s_{46}s_{12}}{s_{34}s_{346}}f^{\text{DT}}(s_{12},s_{25})+\frac{s_{12}s_{25}s_{36}s_{46}}{s_{346}}f^{\text{ST}}(s_{34},s_{36})f^{\text{DT}}(s_{12},s_{25})\\
    &-\kappa^2\frac{s_{45}s_{16}s_{35}}{s_{345}}f^{\text{ST}}(s_{34},s_{45})-g^2\frac{s_{12}s_{35}s_{16}}{s_{34}s_{345}}f^{\text{DT}}(s_{12},s_{26})+\frac{s_{12}s_{16}s_{35}s_{45}}{s_{345}}f^{\text{ST}}(s_{34},s_{45})f^{\text{DT}}(s_{12},s_{26})\\
    &+\frac{N_6[12;3456]}{s_{34}}+CT_6[12;3456]\,.
\end{aligned}
\end{equation}
The first line is the pure $\mathcal N=4$ EYM amplitude; the next two lines supply the higher-derivative contributions to the three-particle residues. The remaining numerator $N_6[12;3456]$ is constrained by factorization at $s_{34}=0$, while $CT_6[12;3456]$ parametrizes local contact terms. Both are initially polynomials in Mandelstam variables and Levi-Civita contractions. The matching condition at $[34]=0$ is
\begin{equation}
\label{eq:N6_<34>matching}
    A[1^{12} 2^{13};3^{14}4^{23} 5^{34} 6^{24}]\bigg|_{[34]\to 0} = -g\left(-\frac{s_{235}}{s_{34}}+\frac{s_{25}s_{46}}{s_{34}s_{346}}+\frac{s_{16}s_{35}}{s_{34}s_{345}}\right)\frac{N_5[12;56P]}{[12]},
\end{equation}
where $P=p_3+p_4$. Exchanging angle and square brackets on the right-hand side gives the matching condition at $\la34\ra=0$.

Following~\cite{Elvang:2026pmc}, we express $N_6$ in terms of the six-point Mandelstam variables
\begin{equation}\label{eq:6point_Mandelstam}
    \{s_{12},s_{23},s_{34},s_{45},s_{56},s_{16}, s_{123},s_{234},s_{345}\}
\end{equation}
and the parity-odd Levi-Civita contractions
\begin{equation}
\begin{aligned}
    &\{\epsilon_{1234},\epsilon_{1235},\epsilon_{1245}, \epsilon_{1345},\epsilon_{2345}\},
\end{aligned}
\end{equation}
where 
\begin{equation}
    \epsilon_{ijkl}=4i\epsilon_{\mu\nu\rho\sigma}p_i^\mu p_j^\nu p_k^\rho p_l^\sigma = [ij]\la jk\ra[kl]\la li\ra - \la ij\ra[jk]\la kl\ra[li].
\end{equation}
Products of two Levi-Civita contractions reduce to Mandelstam polynomials, so the ansatz need only be linear in each contraction. The matching condition~\eqref{eq:N6_<34>matching} and the five-point ansatz~\eqref{eq:N5_ansatz} restrict the parity-odd term in $N_6$ to $\epsilon_{1256}=\epsilon_{1235}+\epsilon_{1245}$, giving
\begin{equation}
    N_6 = N_{6}^{(1)}(s_{ij}) + \epsilon_{1256} N_{6}^{(2)}(s_{ij}).
\end{equation}
Here $N_6^{(1)}$ and $N_6^{(2)}$ are Mandelstam polynomials. On the $s_{34}=0$ pole, they can depend on arbitrary powers of the five-point variables
\begin{equation}\label{eq:N6_five_point_Mandelstam}
    \{s_{12},s_{234},s_{345},s_{56},s_{16}\}
\end{equation}
but are at most linear in
\begin{equation}
    \{s_{123},s_{23},s_{45}\}
\end{equation}
because the $s_{34}$ channel exchanges a gluon. We therefore use
\begin{equation}
    N_{6}^{(i)} = N_{6}^{(i,1)} + s_{123} N_{6}^{(i,2)} + s_{23} N_{6}^{(i,3)} + s_{45} N_{6}^{(i,4)},\quad i=1,2, 
\end{equation}
where $N_6^{(i,j)}$ are polynomials in the variables of~\eqref{eq:N6_five_point_Mandelstam}. Two-particle residue matching fixes all parameters within this ansatz at each order studied, as shown in table~\ref{tab:N6_residue_matching}. 
\begin{table}[h!]
    \centering
    \begin{tabular}{|c|c|c|c|c|c|c|c|c|c|}
    \hline
    Order in $N_6$ & $s^0$ & $s^1$ & $s^2$ & $s^3$ & $s^4$ & $s^5$ & $s^6$ & $s^7$ & $s^8$\\
    \hline
    Ansatz & 1 & 8 & 31 & 88 & 205 & 416 & 763 & 1296 & 2073\\
    \hline
    Matching residue $[34],\la34\ra \to 0$ & 0 & 0 & 0 & 0 & 0 & 0 & 0 & 0 & 0\\
    \hline
    \end{tabular}
    \caption{Number of parameters in the $N_6$ ansatz at each order, before and after two-particle residue matching. All parameters within the ansatz are fixed in terms of four- and five-point coefficients.}\label{tab:N6_residue_matching}
\end{table}

The factorization residues obey the Ward identities because they are sewn from lower-point superamplitudes. Any remaining violation is therefore polynomial and must be canceled by $CT_6[12;3456]$. In the analyses of~\cite{Elvang:2026pmc,Berman:2026ezk}, unrestricted contact terms can absorb these violations. Requiring separate $C$ and $P$ invariance instead imposes the scalar relation
\begin{equation}
A[1^{12}2^{13};3^{14}4^{23}5^{34}6^{24}]=A[1^{34}2^{24};3^{23}4^{14}5^{12}6^{13}]\,,
\end{equation}
and yields nonlinear relations among the four-point Wilson coefficients. In the pole decomposition used here, this restricts $CT_6[12;3456]$ to parity-even polynomials in the Mandelstam variables~\eqref{eq:6point_Mandelstam}.

\subsubsection{The $\tr(T^2)\tr(T^2)\tr(T^2)$ sector}

For the triple-trace sector, we choose $A[1^{12}2^{13};3^{14}4^{34};5^{23}6^{24}]$ and the seven other scalar assignments generated by $(12)$, $(34)$, and $(56)$. This component has poles at $s_{124}=0$, $s_{126}=0$, and $s_{134}=0$. Matching their residues gives 
\begin{equation}
\begin{aligned}
    &A[1^{12}2^{13};3^{14}4^{34};5^{23}6^{24}] = \frac{s_{12}s_{24}s_{36}s_{56}}{s_{124}}\left(\frac{-\kappa^2}{s_{12}}+f^{\text{DT}}(s_{12},s_{24})\right)\left(\frac{-\kappa^2}{s_{56}}+f^{\text{DT}}(s_{56},s_{35})\right)\\
    &+\frac{s_{12}s_{16}s_{34}s_{45}}{s_{126}}\left(\frac{-\kappa^2}{s_{12}}+f^{\text{DT}}(s_{12},s_{26})\right)\left(\frac{-\kappa^2}{s_{34}}+f^{\text{DT}}(s_{34},s_{35})\right)\\
    &+\frac{s_{13}s_{34}s_{25}s_{56}}{s_{134}}\left(\frac{-\kappa^2}{s_{34}}+f^{\text{DT}}(s_{34},s_{14})\right)\left(\frac{-\kappa^2}{s_{56}}+f^{\text{DT}}(s_{56},s_{25})\right) + CT_6[12;34;56],
\end{aligned}
\end{equation}
where $CT_6[12;34;56]$ contains the local contact terms. We again impose scalar parity by taking it to be a general polynomial in the Mandelstam variables~\eqref{eq:6point_Mandelstam}.

\section{Nonlinear SUSY constraints}
\label{sec:constraints}

The Ward identities in the $\tr(T^2)^3$ and $\tr(T^2)\tr(T^4)$ sectors give nonlinear constraints on $b_{k,q}$ alone and on $a_{k,q}$ and $b_{k,q}$ jointly. We derive them through $a_{7,q}$ and $b_{8,q}$; the offset between the two derivative orders reflects the mass dimension of $\kappa$.

\subsection{Two solution branches}
The first mixed-sector constraints, organized by inverse mass dimension, are
\begin{equation}
\begin{aligned}\label{eq:ab_nonlinear_example}
    O(s^{-3}):\quad & \kappa^2 a_{0,0}=0,\\
    O(s^{-4}):\quad & a_{0,0} b_{0,0}=0,\ \kappa ^2 a_{1,0}=4 g^2 b_{2,2},\\
    O(s^{-5}):\quad & \kappa ^2 a_{2,0}= \kappa ^2 a_{2,1}=  a_{0,0} b_{1,0}=0,\ g^2 b_{3,2}=-\frac{1}{4} a_{1,0} b_{0,0},\\
    O(s^{-6}):\quad & a_{2,0} b_{0,0}= a_{2,1} b_{0,0}= a_{0,0} b_{2,0}= a_{0,0} b_{2,2}=0,\\
    &\kappa ^2 a_{3,0}= \frac{1}{2}\kappa ^2 a_{3,1}= 16 g^2 b_{4,4},\ g^2 b_{4,2}=2 g^2 b_{4,4}-\frac{1}{4} a_{1,0} b_{1,0}.\\
\end{aligned}
\end{equation}
The first constraint, $\kappa^2a_{0,0}=0$, separates the two nondegenerate branches $\kappa=0$, $a_{0,0}\neq0$ and $\kappa\neq0$, $a_{0,0}=0$.\footnote{We also assume $g\neq0$.} We call these the \emph{type-I} and \emph{heterotic} branches, respectively. On the heterotic branch, higher-point supersymmetry and scalar parity forbid the single-trace $F^4$ interaction, although it is allowed by four-point supersymmetry. The complete mixed- and triple-trace constraints through $a_{7,q}$ and $b_{8,q}$ are supplied in the ancillary file.\\

\noindent \textbf{Type-I branch}

On the type-I branch, the mixed-sector constraints set $b_{k,q}=0$ through $k=8$, which also solves the triple-trace constraints. We expect this pattern to persist to all orders, reducing the system to the open-superstring constraints of~\cite{Elvang:2026pmc}. Appendix~\ref{sec:single_trace_limit} discusses the relation between vanishing double-trace data and single-trace positivity.\\

\noindent \textbf{Heterotic branch}

On the heterotic branch, we combine the mixed- and triple-trace constraints with the single-trace relations of~\cite{Elvang:2026pmc} through $a_{7,q}$ and $b_{8,q}$. We normalize the coefficients by the corresponding couplings,
\begin{equation}
    \hat{a}_{k,q} = \frac{a_{k,q}}{g^2},\qquad \hat{b}_{k,q} = \frac{b_{k,q}}{\kappa^2}.
\end{equation}
The normalized coefficients are determined by five diagonal parameters, $\{\hat b_{0,0},\hat b_{2,2},\hat b_{4,4},\hat b_{6,6},\hat b_{8,8}\}$. The single-trace solution is
\begin{equation}
\begin{aligned}\label{eq:SUSY_akq}
    k=0:\quad &\hat{a}_{0,0}=0,\\
    k=1:\quad &\hat{a}_{1,0}=4 \hat{b}_{2,2},\\
    k=2:\quad &\hat{a}_{2,0}=\hat{a}_{2,1}=0,\\
    k=3:\quad &\hat{a}_{3,0}=\frac{1}{2}\hat{a}_{3,1}=16 \hat{b}_{4,4},\\
    k=4:\quad &\hat{a}_{4,0}=0,\quad \hat{a}_{4,1}=\frac{1}{2}\hat{a}_{4,2}=-8\hat{b}_{2,2}^2\\
    k=5:\quad &\hat{a}_{5,0}=\frac{1}{3}\hat{a}_{5,1}=\frac{1}{5}\hat{a}_{5,2}=64 \hat{b}_{6,6}\\
    k=6:\quad &\hat{a}_{6,0}=0,\quad \hat{a}_{6,1}=\frac{1}{3}\hat{a}_{6,2}=\frac{1}{4}\hat{a}_{6,3}=-64 \hat{b}_{2,2} \hat{b}_{4,4},\\
    k=7:\quad &\hat{a}_{7,0}=\frac{1}{4}\hat{a}_{7,1}=256 \hat{b}_{8,8},\quad \hat{a}_{7,2}=\frac{32}{3} \hat{b}_{2,2}^3+\frac{7168 \hat{b}_{8,8}}{3}, \quad \hat{a}_{7,3}=32 \hat{b}_{2,2}^3+3584 \hat{b}_{8,8}.
\end{aligned}
\end{equation}
The double-trace solution is
\begin{equation}
\begin{aligned}\label{eq:SUSY_bkq}
    k=1:\quad &\hat{b}_{1,0}=-\hat{b}_{0,0}^2,\\
    k=2:\quad &\hat{b}_{2,0}=\hat{b}_{0,0}^3-\hat{b}_{2,2},\\
    k=3:\quad &\hat{b}_{3,0}=\hat{b}_{0,0} \hat{b}_{2,2}-\hat{b}_{0,0}^4,\quad \hat{b}_{3,2}=-\hat{b}_{0,0} \hat{b}_{2,2},\\
    k=4:\quad &\hat{b}_{4,0}=\hat{b}_{0,0}^5-\hat{b}_{2,2} \hat{b}_{0,0}^2-3 \hat{b}_{4,4},\quad \hat{b}_{4,2}=\hat{b}_{2,2} \hat{b}_{0,0}^2+2 \hat{b}_{4,4},\\
    k=5:\quad &\hat{b}_{5,0}=-\hat{b}_{0,0}^6+\hat{b}_{2,2} \hat{b}_{0,0}^3+3 \hat{b}_{4,4} \hat{b}_{0,0}-\frac{1}{2} \hat{b}_{2,2}^2,\\
    &\hat{b}_{5,2}=-2 \hat{b}_{4,4} \hat{b}_{0,0}+\hat{b}_{2,2}^2-\hat{b}_{0,0}^3 \hat{b}_{2,2},\quad \hat{b}_{5,4}=-\frac{1}{2} \hat{b}_{2,2}^2-\hat{b}_{0,0} \hat{b}_{4,4},\\
    k=6:\quad &\hat{b}_{6,0}=\hat{b}_{0,0}^7-\hat{b}_{2,2} \hat{b}_{0,0}^4-3 \hat{b}_{4,4} \hat{b}_{0,0}^2+\frac{1}{2} \hat{b}_{2,2}^2 \hat{b}_{0,0}-9 \hat{b}_{6,6},\\
    &\hat{b}_{6,2}=\hat{b}_{2,2} \hat{b}_{0,0}^4+2 \hat{b}_{4,4} \hat{b}_{0,0}^2-\hat{b}_{2,2}^2 \hat{b}_{0,0}+3 \hat{b}_{6,6},\quad \hat{b}_{6,4}=\hat{b}_{4,4} \hat{b}_{0,0}^2+\frac{1}{2} \hat{b}_{2,2}^2 \hat{b}_{0,0}+5 \hat{b}_{6,6}\\
    k=7:\quad &\hat{b}_{7,0}=-\hat{b}_{0,0}^8+\hat{b}_{2,2} \hat{b}_{0,0}^5+3 \hat{b}_{4,4} \hat{b}_{0,0}^3-\frac{1}{2} \hat{b}_{2,2}^2 \hat{b}_{0,0}^2+9 \hat{b}_{6,6} \hat{b}_{0,0}-3 \hat{b}_{2,2} \hat{b}_{4,4},\\
    &\hat{b}_{7,2}=-2 \hat{b}_{4,4} \hat{b}_{0,0}^3+\hat{b}_{2,2}^2 \hat{b}_{0,0}^2-3 \hat{b}_{6,6} \hat{b}_{0,0}-\hat{b}_{0,0}^5 \hat{b}_{2,2}+5 \hat{b}_{2,2} \hat{b}_{4,4},\\
    &\hat{b}_{7,4}=-\hat{b}_{4,4} \hat{b}_{0,0}^3-\frac{1}{2} \hat{b}_{2,2}^2 \hat{b}_{0,0}^2-5 \hat{b}_{6,6} \hat{b}_{0,0}-\hat{b}_{2,2} \hat{b}_{4,4},\quad \hat{b}_{7,6}=-\hat{b}_{2,2} \hat{b}_{4,4}-\hat{b}_{0,0} \hat{b}_{6,6},\\
    k=8:\quad &\hat{b}_{8,0}=\hat{b}_{0,0}^9-\hat{b}_{2,2} \hat{b}_{0,0}^6-3 \hat{b}_{4,4} \hat{b}_{0,0}^4+\frac{1}{2} \hat{b}_{2,2}^2 \hat{b}_{0,0}^3-9 \hat{b}_{6,6} \hat{b}_{0,0}^2+3 \hat{b}_{2,2} \hat{b}_{4,4} \hat{b}_{0,0}-\frac{1}{6} \hat{b}_{2,2}^3-\frac{85 \hat{b}_{8,8}}{3},\\
    &\hat{b}_{8,2}=\hat{b}_{2,2} \hat{b}_{0,0}^6+2 \hat{b}_{4,4} \hat{b}_{0,0}^4-\hat{b}_{2,2}^2 \hat{b}_{0,0}^3+3 \hat{b}_{6,6} \hat{b}_{0,0}^2-5 \hat{b}_{2,2} \hat{b}_{4,4} \hat{b}_{0,0}+\frac{1}{2} \hat{b}_{2,2}^3+4 \hat{b}_{8,8},\\
    &\hat{b}_{8,4}=\hat{b}_{4,4} \hat{b}_{0,0}^4+\frac{1}{2} \hat{b}_{2,2}^2 \hat{b}_{0,0}^3+5 \hat{b}_{6,6} \hat{b}_{0,0}^2+\hat{b}_{2,2} \hat{b}_{4,4} \hat{b}_{0,0}-\frac{1}{2} \hat{b}_{2,2}^3+14 \hat{b}_{8,8},\\
    &\hat{b}_{8,6}=\frac{1}{6} \hat{b}_{2,2}^3+\hat{b}_{0,0} \hat{b}_{4,4} \hat{b}_{2,2}+\hat{b}_{0,0}^2 \hat{b}_{6,6}+\frac{28 \hat{b}_{8,8}}{3}.
\end{aligned}
\end{equation}
These relations agree with those obtained by treating the odd zeta values in the heterotic expansion~\eqref{eq:heterotic_trace_functions} as independent parameters.

\subsection{Structure of the heterotic branch}

The dependence on diagonal double-trace coefficients is reproduced by the generating functions
\begin{equation}
\begin{aligned}\label{eq:resum_nonlinear_constraints}
    &F^{\text{ST}}(s,t) = -\frac{g^2}{st} e^E,\quad F^{\text{DT}}(s,t)=-\frac{\kappa^2}{s\,(1+\hat b_{0,0}\,s)}\,e^{E},\\
    &E(s,t,u)=\sum_{r\geq 1}\frac{4^r \hat b_{2r,2r}}{2r+1}(s^{2r+1}+t^{2r+1}+u^{2r+1}),
\end{aligned}
\end{equation}
with free coefficients $\hat b_{2r,2r}$ for $r\geq0$. This form reproduces the solution through $a_{7,q}$ and $b_{8,q}$. The heterotic values are
\begin{equation}
\begin{aligned}
    &\hat{b}_{0,0}=\alpha',
    &\hat b_{2r,2r} = 2^{1-2r}\zeta_{2r+1}\alpha'^{2r+1},\quad r\geq 1.\\
\end{aligned}
\end{equation}

The single-trace contact terms cannot be turned on independently of the double-trace terms: the latter determine the former, but not conversely. In particular, $a_{k,q}$ is independent of $\hat b_{0,0}$, which controls the ``tachyon pole'' of $F^{\mathrm{DT}}$ in~\eqref{eq:resum_nonlinear_constraints}. An alternative parametrization keeps $\hat b_{0,0}$ and uses
\begin{equation}\label{eq:b2k2k_a2k-10}
    \hat b_{2k,2k} = \frac{1}{4^k}\hat a_{2k-1,0}\quad k\geq 1
\end{equation}
to replace the remaining diagonal parameters by $\hat a_{2k-1,0}$.

The relations also give rank-one Hankel matrices with entries linear in the Wilson coefficients. One example is
\begin{equation}\label{eq:H_Hankel}
    H=\begin{pmatrix}
        b_{0,0} & b_{1,0} & b_{2,0}+b_{2,2} & \cdots\\
        b_{1,0} & b_{2,0}+b_{2,2} & b_{3,0}+b_{3,2} & \cdots\\
        b_{2,0}+b_{2,2} & b_{3,0}+b_{3,2} & b_{4,0}+b_{4,2} + b_{4,4} & \cdots\\
        \vdots & \vdots & \vdots & \ddots
    \end{pmatrix},
\end{equation}
with entries
\begin{equation}
	H_{i,j}=\sum_{k=0}^{\lfloor\frac{i+j-2}{2}\rfloor}b_{i+j-2,2k}.
\end{equation}
Its rank follows from the forward limit of~\eqref{eq:resum_nonlinear_constraints},
\begin{equation}
    F^{\text{DT}}(s,0)=-\frac{\kappa^2}{s(1+\hat b_{0,0} s)} = -\frac{\kappa^2}{s}\sum_{k=0}^\infty (-\hat b_{0,0}s)^k=-\frac{\kappa^2}{s}+\sum_{k\geq0}\sum_{\substack{0\leq q\leq k\\ q\ \text{even}}}b_{k,q} s^k.
\end{equation}
Equating the last two expressions gives a geometric sequence for the entries of $H$.

\subsection{The IST amplitude}\label{subsec:IST-solution}

Besides the heterotic string amplitude given in \eqref{eq:heterotic_factor} and \eqref{eq:heterotic_trace_functions}, the IST amplitude also satisfies the nonlinear SUSY constraints on the heterotic branch. The IST amplitudes first appeared as primal solutions of the nonlinear $\mathcal N=8$ SUSY constraints in the supergravity bootstrap~\cite{Berman:2026ezk}. They are meromorphic amplitudes with rational residues and infinite spin support at fixed mass. The corresponding family on the heterotic branch satisfies the $\mathcal N=4$ nonlinear SUSY constraints through the orders studied and has reduced functions
\begin{equation}\label{eq:infinite_spin_tower_trace_functions}
F_{\mathrm{IST}}^{\mathrm{ST}}(s,t)
=-\frac{g^2}{s t} W^q(s,t,u),\qquad
F_{\mathrm{IST}}^{\mathrm{DT}}(s,t)
=-\frac{\kappa^2}{s(1+s/m^2)} W^q(s,t,u).
\end{equation}
Here $q$ is a positive integer and
\begin{equation}\label{eq: WFunction}
    W^q(s,t,u) =\prod_{n=1}^q \frac{(nm^2+s)(nm^2+t)(nm^2+u)}{(nm^2-s)(nm^2-t)(nm^2-u)}.
\end{equation}
The massive poles occur at $s=nm^2$, $n=1,\ldots,q$, and in the crossed channels, so $m=M_{\rm gap}$ is the physical mass of the first massive level. The dispersive cutoff can be chosen anywhere in $0<\Lambda\leq m$. The limit $q\to\infty$ gives the heterotic amplitude with $\alpha'=m^{-2}$ \footnote{Reference~\cite{Berman:2026ezk} also considers analytic continuation to real $q$.}. Its Wilson coefficients follow by replacing the zeta values in the heterotic expansion with generalized harmonic numbers,
\begin{equation}
\zeta_k=\sum_{n=1}^\infty \frac{1}{n^k} \quad \longrightarrow \quad H_q^k=\sum_{n=1}^q \frac{1}{n^k}\,.
\end{equation}
For $q=1$, the first coefficients are
\begin{equation}
\begin{aligned}
    &a_{0,0}=0,\quad &a_{1,0}&=2 \frac{g^2}{m^6},\quad &a_{2,0}&=0,\\
    &a_{2,1}=0,\quad &a_{3,0}&=2 \frac{g^2}{m^{10}},\quad &a_{3,1}&=4 \frac{g^2}{m^{10}},\\
    &a_{4,0}=0,\quad &a_{4,1}&=-2 \frac{g^2}{m^{12}},\quad &a_{4,2}&=-4 \frac{g^2}{m^{12}},\\
    &b_{0,0}=\frac{\kappa ^2}{m^2},\quad &b_{1,0}&=-\frac{\kappa ^2}{m^4},\quad &b_{2,0}&=\frac{\kappa ^2}{2m^6},\\
    &b_{2,2}=\frac{\kappa ^2}{2m^6},\quad &b_{3,0}&=-\frac{\kappa ^2}{2m^8},\quad &b_{3,2}&=-\frac{\kappa ^2}{2m^8}.
\end{aligned}
\end{equation}
For integer $N\geq4$ and nonzero gauge coupling, partial-wave unitarity gives\footnote{The abelian $SO(2)$ case and the degeneracies of the projector basis at $SO(3)$ require separate treatment and are not included in these bounds.}
\begin{equation}
\begin{aligned}
    &4\leq N \leq 7:\quad &0< \frac{g^2}{\kappa^2m^2} \leq 4,\\
    &8\leq N \leq 48: &\frac{3 N(N{-}1){-}(4 N(N{-}1)+16)\log (2)}{(4 N{-}8) \log (2)} \leq \frac{g^2}{\kappa^2m^2} \leq 4\,
\end{aligned}
\end{equation}
as derived in appendix~\ref{sec:Unitarity_IST}; no unitary solution exists for $N\geq49$.

\section{Bootstrapping the heterotic string}\label{sec:branches}
We now impose unitarity through the partial-wave expansion in angular momentum and irreducible color representations, then combine the resulting dispersive constraints with the nonlinear relations.

\subsection{Dispersion relations in color space} \label{sec:dispersion}
For the scalar component amplitude $A(1^{12}2^{12}3^{34}4^{34}) = s^2 A_4(s,t)$ where the scalar are adjoints of $SO(N)$, we expand the amplitude in the orthogonal projector basis
\begin{equation}\label{eq:partial_wave_expansion}
A(1^{12}2^{12}3^{34}4^{34})=\sum_{J,R} a_{J,R}(s) \mathbb{P}^R_s P_J(1{+}\frac{2t}{s}),
\end{equation}
where $a_{J,R}$ are partial-wave coefficients, $P_J$ are Legendre polynomials, and $\mathbb P_s^R$ project onto the irreducible representations exchanged in the $s$ channel,
\begin{equation}
    \mathbf{Adj}\otimes \mathbf{Adj}=\bigoplus_{R} R.
\end{equation}
Partial-wave unitarity requires
\begin{equation}
    \text{Im}\, a_{J,R}(s)\geq0,\quad s>0\,,
\end{equation}
for every $R$ and $J$.

The six projectors for adjoint $SO(N)$ external states are related linearly to the six trace structures,
\begin{equation}
\vec{\mathbb{P}}_s=M_s \vec{\mathbb{T}},
\end{equation}
where 
\begin{equation}\label{eq:Ms_trace_matrix}
M_s =
\begin{pmatrix}
0 & 0 & 0 & \dfrac{2}{N(N{-}1)} & 0 & 0 \\
\dfrac{2}{N{-}2} & \dfrac{2}{N{-}2} & 0 & {-}\dfrac{4}{N(N{-}2)} & 0 & 0 \\
{-}\dfrac{2}{N{-}2} & {-}\dfrac{2}{N{-}2} & \dfrac{2}{3} & \dfrac{2}{(N{-}1)(N{-}2)} & \dfrac{1}{3} & \dfrac{1}{3} \\
0 & 0 & -\dfrac{2}{3} & 0 & \dfrac{1}{6} & \dfrac{1}{6} \\
\dfrac{2}{N{-}2} & -\dfrac{2}{N-2} & 0 & 0 & 0 & 0 \\
-\dfrac{2}{N-2} & \dfrac{2}{N-2} & 0 & 0 & -\dfrac{1}{2} & \dfrac{1}{2}
\end{pmatrix},
\qquad
\vec{\mathbb T}=
\begin{pmatrix}
\tr(1,2,3,4) \\
\tr(1,3,4,2) \\
\tr(1,4,2,3) \\
\tr(1,2)\tr(3,4) \\
\tr(1,3)\tr(2,4) \\
\tr(1,4)\tr(2,3)
\end{pmatrix}\,,
\end{equation}
where $N\geq4$. We order the representations so that $R=1,2,3,4$ are symmetric and $R=5,6$ antisymmetric. The singlet is $R=1$, the symmetric traceless representation is $R=2$, and the adjoint is $R=5$. Bose symmetry restricts the first four to even spin and the last two to odd spin.

In fixed-$t$ dispersion relations, the $s$- and $u$-channel discontinuities have positive spectral densities in their respective projector bases. The bases are related by
\begin{equation}
    \vec{\mathbb P}_s = M_{su} \vec{\mathbb P}_u,
\end{equation}
where
\begin{equation}
    M_{su}=\begin{pmatrix}
    \frac{2}{(N-1) N} & \frac{2}{(N-1) N} & \frac{2}{(N-1) N} & \frac{2}{(N-1) N} & -\frac{2}{(N-1) N} & -\frac{2}{(N-1) N} \\
    \frac{N+2}{N} & \frac{N^2-8}{2 (N-2) N} & \frac{N-4}{(N-2) N} & -\frac{2 (N+2)}{(N-2) N} & -\frac{(N-4) (N+2)}{2 (N-2) N} & \frac{4}{(N-2) N} \\
    \frac{(N-3) (N+1) (N+2)}{6 (N-1)} & \frac{(N-4) (N-3) (N+1)}{6 (N-2) (N-1)} & \frac{N^2-6 N+11}{3 (N-2) (N-1)} & \frac{(N+1) (N+2)}{3 (N-2) (N-1)} & \frac{(N-3) (N+1) (N+2)}{6 (N-2) (N-1)} & \frac{(N-4) (N+1)}{3 (N-2) (N-1)} \\
    \frac{1}{12} (N-3) (N-2) & \frac{3-N}{6} & \frac{1}{6} & \frac{1}{6} & \frac{3-N}{6} & \frac{1}{6} \\
    -1 & -\frac{N-4}{2 (N-2)} & \frac{1}{N-2} & -\frac{2}{N-2} & \frac{1}{2} & 0 \\
    -\frac{1}{4} (N-3) (N+2) & \frac{N-3}{N-2} & \frac{N-4}{2 (N-2)} & \frac{N+2}{2 (N-2)} & 0 & \frac{1}{2} 
    \end{pmatrix}.
\end{equation}

We assume that the superamplitude grows more slowly than $s^2$ at large $s$ and fixed $t$,
\begin{equation}
    \lim_{s\to \infty}\frac{\mathcal{A}[G^4](s,t)}{s^2}=0.
\end{equation}
Factoring out the supermomentum delta function, which scales as $s^2$, gives a reduced function that vanishes in this limit. Hence
\begin{equation}
    \oint_{\infty} \frac{d s'}{2\pi i (s'-s)}A_4(s',t) =0,
\end{equation}
with $A_4$ defined in~\eqref{eq:single_double_tr}. Since the color traces are kinematics-independent, the contour integral at infinity vanishes separately for each reduced trace function. Deforming the contour gives
\begin{equation}
\begin{aligned}\label{eq:dispersion_0_subtraction}
    &(\text{Res}_{s'=0}+\text{Res}_{s'=s}+\text{Res}_{s'=-t}) \left[\frac{A_4(s',t)}{s'-s}\right]\\
    &=\sum_{J,R}\int_{\Lambda^2}^\infty ds'\rho_{J,R}(s')\left(\frac{\mathbb{P}^R_s}{s'-s}+\frac{\mathbb{P}^R_u}{s'+s+t}\right)P_{J}\left(1+\frac{2t}{s'}\right).
\end{aligned}
\end{equation}
The spectral density is $\rho_{J,R}(s')=\operatorname{Im}a_{J,R}(s')/(\pi s'^2)$, with the two powers of $s'$ arising from the supermomentum delta function.

Expanding~\eqref{eq:dispersion_0_subtraction} around $t=0$ gives dispersive representations for the Wilson coefficients, except in the $\tr(14)\tr(23)$ sector. There, $F^{\mathrm{DT}}(t,s)$ has a $t$-channel pole on the left-hand side that is not manifest term by term on the right-hand side, preventing coefficient matching. We instead use the once-subtracted relation
\begin{equation}
\begin{aligned}\label{eq:dispersion_1_subtraction}
    &(\text{Res}_{s'=0}+\text{Res}_{s'=s}+\text{Res}_{s'=-t}) \left[\frac{F^{\text{DT}}(t,s')(2s'+t)}{(s'-s)(s'+t)s'}\right]\\
    &=\left[\sum_{J,R}\int_{\Lambda^2}^\infty ds'\rho_{J,R}(s')\frac{(2s'+t)}{(s'+t)s'}\left(\frac{\mathbb{P}^R_s}{s'-s}-\frac{\mathbb{P}^{R}_u}{s'+s+t}\right)P_{J}\left(1+\frac{2t}{s'}\right)\right]_{\tr(14)\tr(23)},
\end{aligned}
\end{equation}
where $[\cdots]_{\tr(14)\tr(23)}$ extracts the indicated trace component.

Using~\eqref{eq:dispersion_1_subtraction} for $\tr(14)\tr(23)$ and~\eqref{eq:dispersion_0_subtraction} for the other trace structures, we obtain dispersive representations for the dimensionless coefficients $a_{k,q}\Lambda^{2k+4}$ and $b_{k,q}\Lambda^{2k+4}$, together with linear null constraints from crossing symmetry.

\subsection{Relaxation of nonlinear constraints}
\label{subsec:bootstrap_setup}

We have now seen that the bootstrap problem consists of the usual color dispersive setup, supplemented by additional constraints arising from consistent six-point factorization and the SUSY Ward identities. These nonlinear constraints turn out to be extremely powerful, in some cases restricting the allowed theory space to a narrow strip containing the heterotic string.

The main conceptual challenge, therefore, is to incorporate the nonlinear constraints as fully as possible. While the dispersive bootstrap has been studied extensively in works such as \cite{Caron-Huot:2020cmc}, the additional constraints encountered here are generically nonlinear, and in many cases non-convex, in the Wilson coefficients. They therefore cannot be imposed directly within the standard semidefinite approach.

Our strategy is to construct a hierarchy of increasingly refined relaxations of the exact nonlinear problem. The key observation is that, once a small number of low-energy parameters are treated as external parameters and assigned fixed numerical values, the nonlinear constraints are linearized. At each fixed choice of these parameters, the resulting problem can then be analyzed using standard convex optimization techniques. We choose the external parameters to be
\begin{equation}\label{eq:external_param}
	x = \Lambda^2 \frac{b_{00}}{\kappa^2},
	\qquad
	u = \Lambda^6 \frac{b_{22}}{\kappa^2},
	\qquad
	r = \Lambda^{-2} \frac{g^2}{\kappa^2}.
\end{equation}
Here, $\Lambda$ is the lower bound of the dispersive integral. Note that in our numeric setup, it need not coincide with the scale where the \emph{actual} physical state first appears, $M_{gap}$. Therefore, there is a rescaling ambiguity in the parameters above: for a fixed spectral density $\rho_{J,R}(m^2)$, one could always generate a family of parameters by tuning the EFT cutoff:
\begin{equation}
	0 < \frac{\Lambda}{M_{\text{gap}}} \leq 1.
\end{equation}
A fixed physical theory, therefore, traces a curve in the space of $(x,u,r)$. These parameters therefore characterize a theory together with a choice of cutoff. From now on, without loss of generality, we shall measure all masses in units of the cutoff, so that $\Lambda = 1$.

With our choice of external parameters, \eqref{eq:external_param}, the nonlinear constraints collected in \eqref{eq:SUSY_akq} and \eqref{eq:SUSY_bkq} naturally organize themselves into three groups:

\begin{itemize}
	\item \textbf{Group 1: linear at fixed $x$}

	The constraints among the $a$ coefficients include
	\begin{align*}
		a_{00}=a_{20}=a_{21}=a_{40}=a_{31}-2a_{30}=a_{42}-2a_{41}=\cdots=0,
	\end{align*}
	while those among the $b$ coefficients include
	\begin{align*}
		b_{10} &= -x\,b_{00}, 
		&b_{20}& = x^2 b_{00}-b_{22},\\
		b_{30} &= -x^3 b_{00}+x\,b_{22}, 
		&b_{32}& = -x\,b_{22},\\
		b_{40} &= x^4 b_{00}-x^2b_{22}-3b_{44}, 
		&b_{42}& = x^2b_{22}+2b_{44},\\
		&\cdots
	\end{align*}
    and more. The full set can be found in the ancillary file.

	\item \textbf{Group 2: linear at fixed $u$}

	\begin{align*}
		b_{52}+x^3b_{22}+2x\,b_{44}&=u\,b_{22},\\
		-b_{76}-x\,b_{66}&=u\,b_{44},\\
		6b_{86}+6x\,b_{76}-56b_{88}
		&=u\,(b_{52}+x^3b_{22}+2x\,b_{44})
	\end{align*}

	\item \textbf{Group 3: linear at fixed $r$}

	\begin{align*}
		a_{10}&=4r\,b_{22},
		&
		a_{41}&=-8r\,(b_{52}+x^3b_{22}+2x\,b_{44}),
		\\
		a_{30}&=16r\,b_{44},
		&
		a_{61}&=64r\,(b_{76}+x\,b_{66}),
		\\
		a_{50}&=64r\,b_{66},
		&
		3a_{72}-28a_{70}
		&=32r\,(6b_{86}+6x\,b_{76}-56b_{88}),
		\\
		a_{70}&=256r\,b_{88}
	\end{align*}
\end{itemize}
In particular, this organization suggests a natural sequence of bootstrap problems: to exploit the constraints in Groups 1, 2, and 3, we scan over the corresponding external parameters that make them linear, while relaxing the remaining nonlinear constraints. Increasing the scan dimension then systematically improves the approximation to the full nonlinear problem. 

\subsection{Numerical implementation}

The organization of the nonlinear constraints above suggests a natural sequence of bootstrap problems. In particular, we consider the following:
\begin{enumerate}
	\item at fixed $x$, upper and lower bounds on $u=\frac{b_{22}}{\kappa^2}$,
	\item at fixed $(x,u)$, upper and lower bounds on $r=\frac{g^2}{\kappa^2}$,
	\item at fixed $(x,u)$, test whether the point is feasible.
\end{enumerate}

\paragraph{Fixed $x$, optimize $u$} For the first problem, we impose all constraints in Group 1 and supplement them with a simple convex relaxation of some relations in Groups 2 and 3. To see the origin of this relaxation, write
$S=b_{52}+x^3b_{22}+2xb_{44}$.
Using $b_{22}=\kappa^2u$ and $g^2=\kappa^2r$, the exact
relations in Groups 2 and 3 give
\begin{equation}\label{eq:HA_HB_SUSY_origin}
    S=u b_{22}=\kappa^2u^2,\qquad
    \frac{a_{10}}{4}=r b_{22}=g^2u,\qquad
    -\frac{a_{41}}{8}=rS=g^2u^2.
\end{equation}
These relations can be repackaged into the following positive-semidefinite matrices,
\begin{eqnarray}\label{eq:HA_HB_factorization}
    H_A \equiv
	\begin{pmatrix}
		g^2 & a_{10}/4 \\
		a_{10}/4 & -a_{41}/8
	\end{pmatrix}=g^2vv^{\mathsf T}\succeq 0,\quad
    H_B\equiv
	\begin{pmatrix}
		\kappa^2 & b_{22} \\
		b_{22} & b_{52}+x^3 b_{22}+2x b_{44}
	\end{pmatrix}=\kappa^2vv^{\mathsf T}\succeq 0,\nonumber\\
\end{eqnarray}
where $ v=\begin{pmatrix}1\\u\end{pmatrix}$. For nonzero couplings, both matrices are positive semidefinite
and rank one, and share the same vector $v$.
At fixed $x$, their entries are linear in the remaining
optimization variables. We obtain a convex relaxation by
retaining only the separate positive-semidefinite conditions.

Let us briefly describe how bounds can be computed. Once the external parameters are fixed, the equality constraints are affine in the couplings and spectral densities, and the feasible set is determined by
\begin{align}
	\vec c_\kappa \kappa^2
	+\vec c_g g^2
	+\sum_{J,R}\int_1^\infty ds'\,
	\rho_{J,R}(s')\,\vec N_{J,R}(s')
	= 0, \nonumber\\
	H_A, H_B \succeq 0, \quad g^2,\kappa^2,\rho_{J,R}(s') \geq 0, 
\end{align}
where $\vec c_\kappa$ and $\vec c_g$ have non-zero components whenever the couplings appear in the constraint, and $\vec{N}$ denotes the collection of constraint kernels, all depending on the external parameters over which we scan. Bounding a ratio of two positive linear functionals is then a standard linear-fractional conic program, whose dual problem may be written as
\begin{align}
	\underset{\beta,\vec\alpha,Y_A,Y_B}{\operatorname{Minimize}}
	\quad & \beta \nonumber\\
	\text{subject to}\quad
	&
	\vec\alpha\cdot\vec N_{J,R}(s')
	-B_{J,R}(s')
	-\tr\!\left[Y_A M^A_{J,R}(s')\right]
	-\tr\!\left[Y_B M^B_{J,R}(s')\right]
	\geq0, \,\,
	\forall\,J,R, \text{ and } s'\ge1,
	\nonumber 
    \\
	&
	\beta+\vec\alpha\cdot\vec c_\kappa
	-\tr(Y_B E_{11})=0, 
    \quad
	\vec\alpha\cdot\vec c_g
	-\tr(Y_A E_{11})=0,
	\qquad
	Y_A,Y_B\succeq0,
\end{align}
where $B_{J,R}$ is the dispersive kernel of $b_{22}$, and $E_{11}=\operatorname{diag}(1,0)$. This is a polynomial program that can be solved using \texttt{SDPB} \cite{Simmons-Duffin:2015qma,Landry:2019qug}. One could easily verify that every dual feasible solution gives a rigorous upper bound on $u$:
\begin{equation}
	\beta\kappa^2-b_{22}
	\geq
	\tr(Y_AH_A)+\tr(Y_BH_B)
	\geq0
	\quad 
	\Rightarrow
	\quad
	u\leq\beta.
\end{equation}
Lower bounds follow in the same way. 

\paragraph{Fixed $(x,u)$, optimize $r$} Scanning over both $x$ and $u$ means that we can impose the rank-1 constraints exactly: at fixed $(x,u)$, the complete constraints in Groups 1 and 2 can be imposed, and we further impose constraints in Group 3 that become affine after eliminating $r$, which include:
\begin{equation}
	a_{10}=4u\,g^2,\qquad
	a_{41}=-8u^2g^2,\qquad
	a_{61}=-4u\,a_{30},\qquad
	3a_{72}-28a_{70}=32u^3g^2.
\end{equation}
The same logic as above, now without the PSD relaxation, can then be used to maximize or minimize the ratio
\begin{equation}
	r=\frac{g^2}{\kappa^2}
	=\frac{a_{10}}{4b_{22}}.
\end{equation}

\paragraph{Fixed ($x$,$u$), feasibility problem}
Finally, we can ask directly whether a given fixed set of external parameters is feasible. The simplest yet non-rigorous way to answer this is to directly minimize $r$ using the setup above. Since it admits the trivial bound $r\geq0$, one expects a feasible point to always return a finite non-negative lower bound. Numerically, points outside the allowed region behave very differently: the lower-bound objective grows toward $+\infty$, while the corresponding upper-bound objective runs toward $-\infty$, until \texttt{SDPB} returns \texttt{maxComplementarity exceeded}. We observe this behavior consistently, and use it as a numerical indicator of infeasibility in our preliminary scans. To obtain a \emph{rigorous} certificate, we instead search for a separating functional. However, let us comment that the bootstrap problem in this work is homogeneous under the overall rescaling
\begin{equation}
	\left\{\kappa^2,g^2,\rho_{J,R}\right\}
	\longrightarrow
	\lambda
	\left\{\kappa^2,g^2,\rho_{J,R}\right\},
	\qquad \lambda>0,
\end{equation}
which allows us to fix $\kappa^2=1$ without loss of generality. We then search for a separating functional, $\vec\alpha$, that satisfies
\begin{equation}
	\vec\alpha\cdot\vec c_\kappa=1,
	\qquad
	\vec\alpha\cdot\vec c_g\geq0,
	\qquad
	\vec\alpha\cdot\vec N_{J,R}(s')\geq0
	\quad\forall\,J,R,\ s'\geq1, 
\end{equation}
whose existence would imply
\begin{equation}
	0
	=
	1+(\vec\alpha\cdot\vec c_g)g^2
	+\sum_{J,R}\int_1^\infty ds'\,
	\rho_{J,R}(s')\,
	\vec\alpha\cdot\vec N_{J,R}(s')
	\geq1,
\end{equation}
a contradiction. Any functional satisfying the conditions above therefore provides a certificate of infeasibility.\footnote{
	A small numerical subtlety arises in the implementation: the leading large-$s'$ coefficients of the polynomials can be summed to zero with positive weight, forcing some coefficients to exactly vanish for every functional, meaning that the semidefinite program is actually of higher co-dimension. We need to solve these zero constraints exactly and reduce the functional dimension before using \texttt{SDPB}, to prevent numerical instability.
}

\paragraph{The remaining constraints} Some nonlinear constraints in Group 3, including 
\begin{equation}
	a_{30}=16r\,b_{44},
	\qquad
	a_{50}=64r\,b_{66},
	\qquad
	a_{70}=256r\,b_{88},
\end{equation}
become affine only after fixing $r$. A further scan over $r$ would therefore impose the complete set of nonlinear constraints. Standard methods utilized in high-dimensional scans in the conformal bootstrap, such as Delaunay triangulation and the cutting-surface algorithms, have been remarkably successful in resolving small bootstrap islands. In our problem, however, the allowed region turns out to be highly non-convex and develops an extremely thin, needle-like shape that is difficult to resolve reliably with adaptive sampling. We leave this three-dimensional scan for future work. However, as we will soon see, despite being only outer approximations of the true allowed region, the bounds are already extremely constraining, as we discuss in the next section.

\section{Results}\label{sec:results}

We use crossing constraints through $k=8$, nonlinear SUSY constraints through $a_{7,q}$ and $b_{8,q}$, and a spin truncation $J_{\max}=100$, following the setup of section~\ref{subsec:bootstrap_setup}.

\begin{figure}[htpb]
	\centering
	\includegraphics[width=0.85\linewidth]{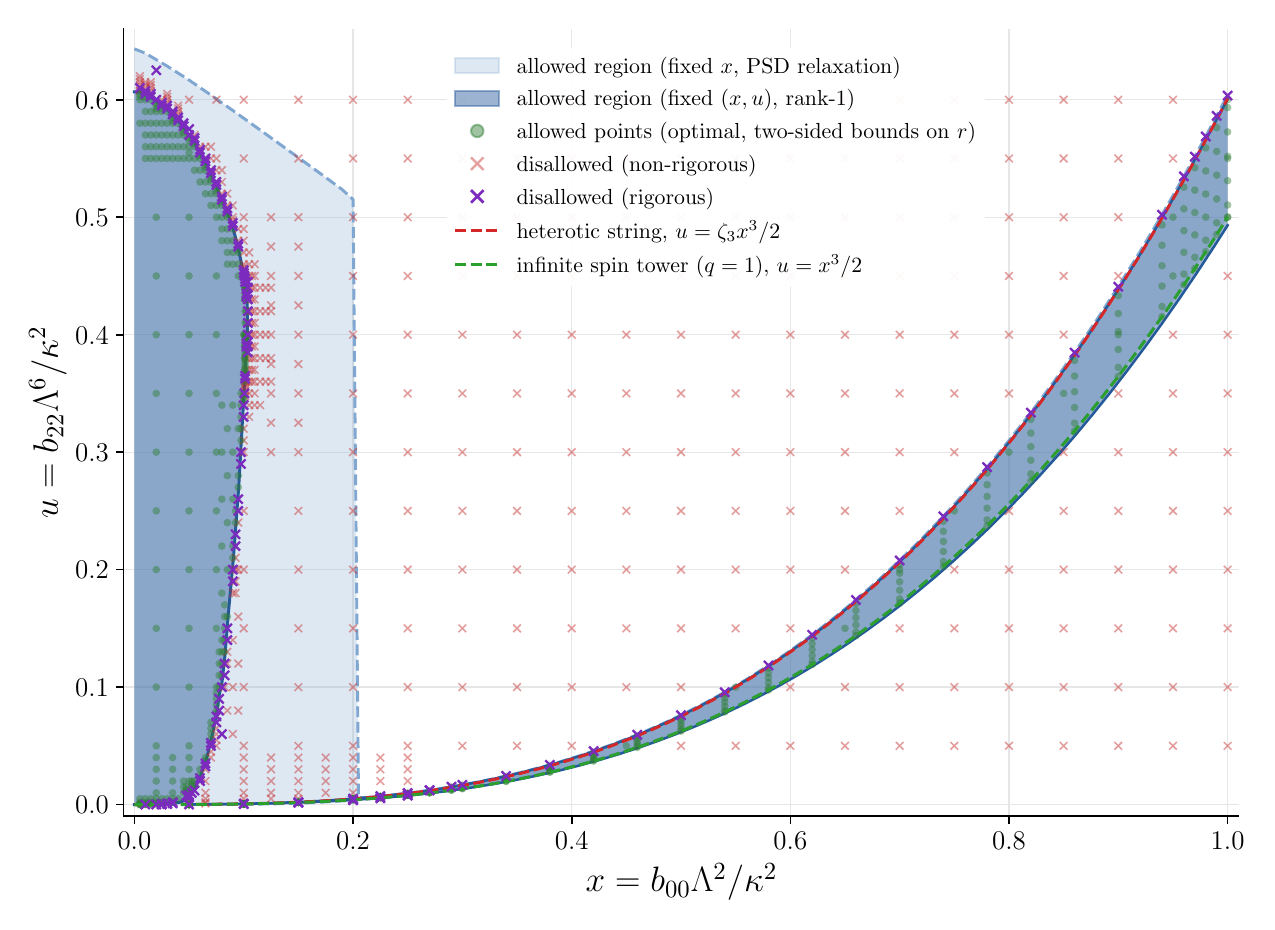}
	\caption{
		Allowed region in $(x,u) = (b_{00}\Lambda^2/\kappa^2,b_{22}\Lambda^6/\kappa^2)$ at $N=10$, focusing on $b_{00}\geq0$. The bounds were obtained at $k_{\max}=8$ and $J_{\max} = 100$. The light blue region shows the PSD relaxation bounds at fixed $x$, while the darker blue region shows the estimated feasible region after imposing the fixed $(x,u)$ rank-one constraints. Green points indicate optimal solutions with two-sided bounds on $r$; shaded boundaries are numerical interpolations. Red points reached the solver’s maximum-complementarity threshold, strongly suggesting while not rigorously establishing infeasibility. Purple points mark rigorous exclusions up to the truncation level, used as a final refinement, which all give feasibility consistent with the non-rigorous results. Trajectories of the heterotic string and $q=1$ IST amplitude closely approach the boundaries of the band on the right.
	}
	\label{fig:b00_b22}
\end{figure}

The allowed region depends non-trivially on $N$ and in general has two parts. Consider the $N=10$ plot in Figure~\ref{fig:b00_b22}. On the right is a narrow band closely bounded by the heterotic amplitude and the $q=1$ IST amplitude defined in~\eqref{eq:infinite_spin_tower_trace_functions}\footnote{For finite $q$ IST, $F^{\mathrm{DT}}(t,s)$ in the $\tr(14)\tr(23)$ sector is $O(s^0)$ at large $s$ and fixed $t$. However, we use the once-subtracted relation~\eqref{eq:dispersion_1_subtraction} for this sector.}. Each generates a cubic trajectory from its endpoint to the origin as the cutoff varies over
\[
	0< \frac{\Lambda}{M_{\text{gap}}}\leq 1.
\]
The heterotic string amplitude generates the trajectory $\hat b_{2,2}=\zeta_3\hat b_{0,0}^3/2$ while the $q=1$ IST generates $\hat b_{2,2}=\hat b_{0,0}^3/2$. These two trajectories lie closely to the upper and lower boundaries of the band in Figure~\ref{fig:b00_b22}. On the left, a broader region near the vertical axis narrows sharply beyond a threshold in $b_{00}$ and joins the band.

\begin{figure}[h!]
    \centering
    \includegraphics[width=0.85\linewidth]{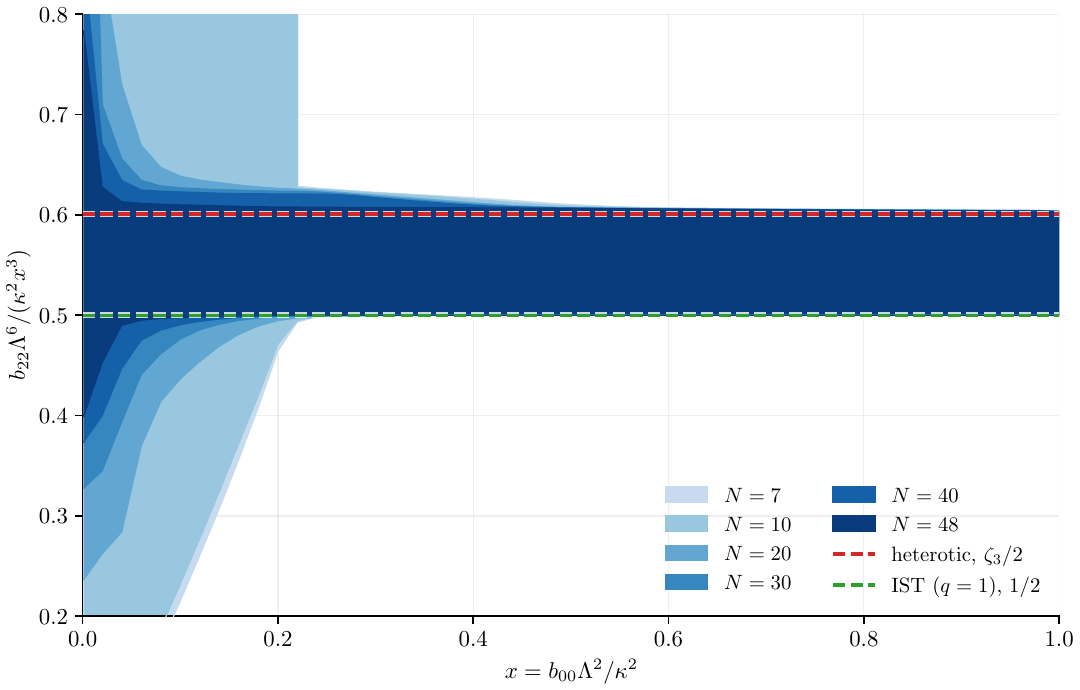}
    \caption{Allowed regions for $N=7,10,20,30,40,48$, shown in increasingly dark shades. With the plotted normalization, the heterotic and $q=1$ IST trajectories become horizontal lines that closely approach the boundaries of the band.}
    \label{fig:b00_b22_N}
\end{figure}

\begin{figure}[htpb]
    \centering
    \includegraphics[width=0.85\linewidth]{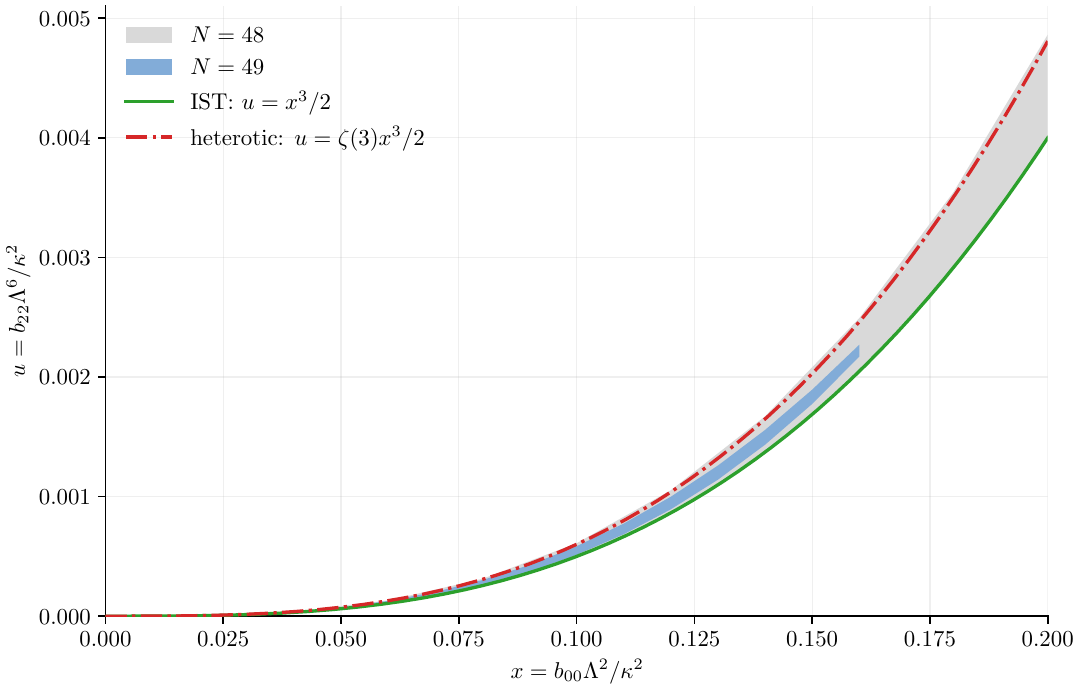}
    \caption{
        Allowed region at $N=49$ (blue), restricted to $x\lesssim0.165$. The band that spans from $x=0$ to $x=1$ at $N=48$ is absent. This is consistent with the fact that neither the heterotic string amplitude nor the $q=1$ IST solution is unitary for $N>48$.
    }
    \label{fig:b00_b22_N49}
\end{figure}

As $N$ increases, the band remains relatively stable while the region near the origin shrinks, becoming small at $N=48$. In Figure~\ref{fig:b00_b22_N}, we illustrate this observation by normalizing  $\frac{b_{2,2}}{\kappa^2}$ against $x^3$. As one can see, the broader region near the vertical axis shrinks dramatically as we increase $N$.
At $N=49$, the band collapses to a small region $x\lesssim0.165$, as shown in Figure~\ref{fig:b00_b22_N49}. This is consistent with the analysis that the heterotic string amplitude and the $q=1$ IST amplitude are non-unitary for $N\geq 49$.

\begin{figure}[htpb]
    \centering
    \includegraphics[width=0.85\linewidth]{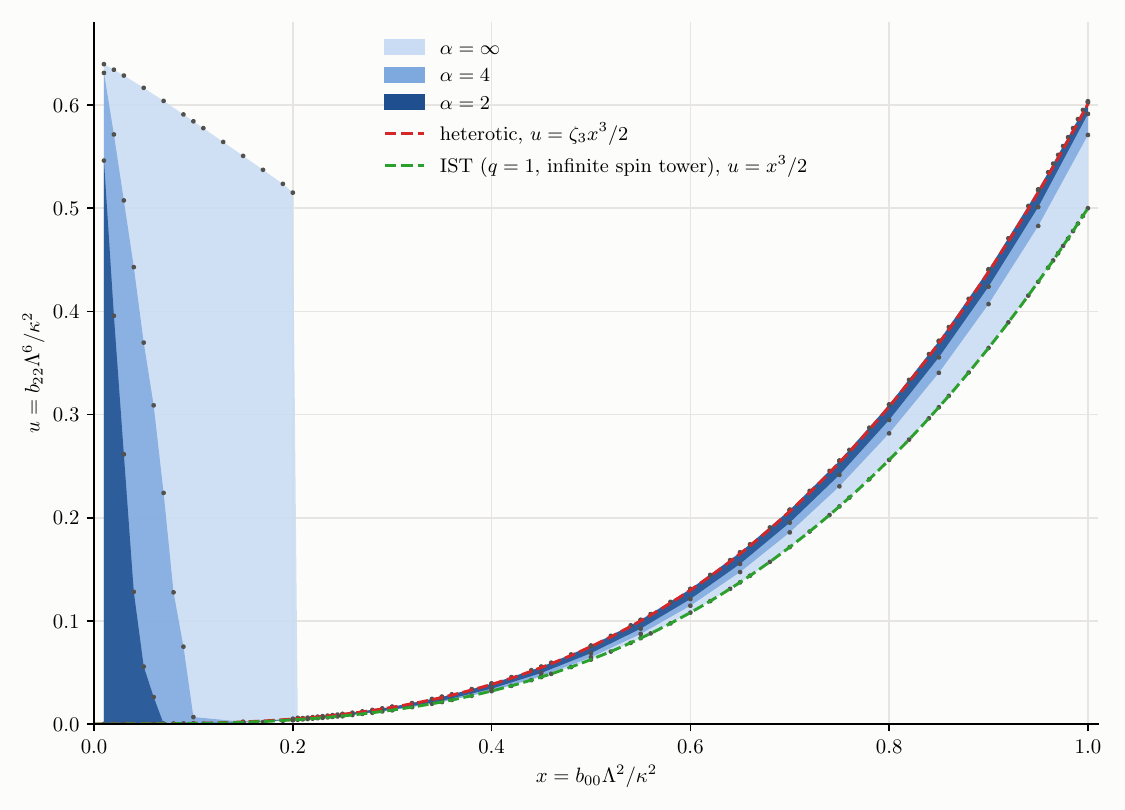}
    \caption{Allowed regions under the spin-dependent gap~\eqref{eq:gap_assumption}, labeled by $\alpha$. Finite $\alpha$ removes much of the broader region near the origin. At $\alpha=2$, an extremely narrow band remains around the heterotic trajectory.}
    \label{fig:b00_b22_slope}
\end{figure}

Motivated by recent analytic constraints relating spin and mass in weakly coupled theories, in particular the sequential spin constraints~\cite{Berman:2024kdh}, we study the region by imposing a spin-dependent gap, a strategy also taken in~\cite{Albert:2024yap,Albert:2022oes,Albert:2023seb}, as follows:
\begin{equation}\label{eq:gap_assumption}
	J \leq 2 + \alpha \left(\frac{m^2}{\Lambda^2} - 1 \right), 
\end{equation}
which defines the support of $\rho_{JR}(m^2)$, where we demand $\rho_{JR}(m^2)=0$ if $J$ violates \eqref{eq:gap_assumption}. These gap assumptions interpolate between $\alpha=2$, saturated by the heterotic string, and $\alpha=\infty$, corresponding to the IST spectrum. A finite $\alpha$ removes much of the region near the origin and narrows the band around the heterotic trajectory, as shown in Figure~\ref{fig:b00_b22_slope}.

\begin{figure}[htpb]
    \centering
    \includegraphics[width=0.85\linewidth]{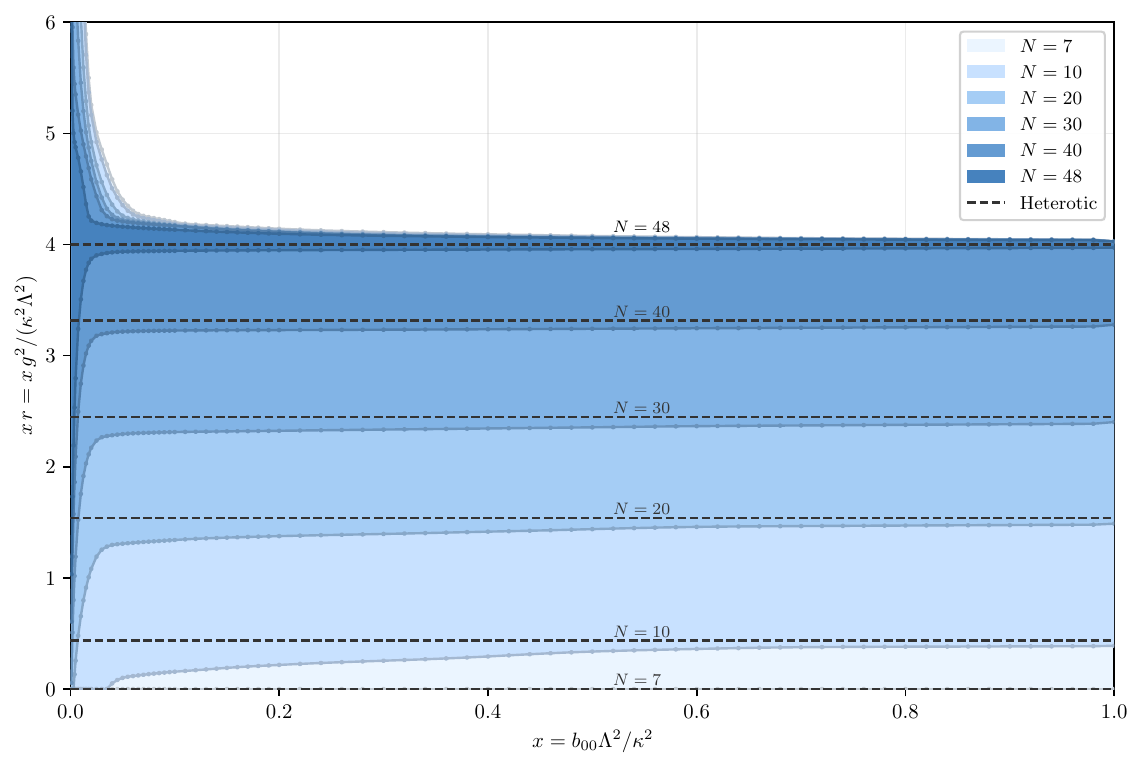}
    \caption{Bounds on the gauge-to-gravity coupling ratio along the heterotic trajectory for different $N$. The plotted quantity $xr$ is invariant under the mass rescaling of section~\ref{subsec:bootstrap_setup}, since $r$ scales as $1/x$. Its lower bounds closely follow the heterotic unitarity limits~\eqref{eq:het_unitary_r}. At $N=48$, the allowed interval is narrow and contains the heterotic value.}
    \label{fig:r_het}
\end{figure}

On the heterotic trajectory, $u=\zeta(3)x^3/2$, we bound $r=g^2 / \kappa^2$ at fixed $(x,u)$, shown in Figure~\ref{fig:r_het}. From the definition, \eqref{eq:external_param}, we see that $r$ has the same scaling as $x^{-1}$ as we tune the ratio $\Lambda/M_{\text{gap}}$, motivating the normalization $xr$. The ratio has a finite upper bound slightly above four in most regions away from $x=0$. Its lower bound closely follows the heterotic unitarity bound
\begin{equation}\label{eq:het_unitary_r}
	\begin{cases}
		4 \leq N \leq 7, & 0 < x r \leq 4, \\
		8 \leq N \leq 48, & \dfrac{N^2 - N - 48}{12N - 24} \leq x r \leq 4.
	\end{cases}
\end{equation}
At $N=48$, we again see that the allowed values lie in a narrow interval containing the heterotic value $xr=4$. These bounds from unitarity is consistent with the analysis of the explicit heterotic amplitude~\cite{Bachu:2022gof}.

\section{Conclusions and outlook}\label{sec:conclusions}
Six-point supersymmetry and peculiar parity impose nonlinear relations between the single- and double-trace Wilson coefficients of four-dimensional $\mathcal N=4$ EYM theory. On the gravitational branch, these relations forbid the single-trace $F^4$ interaction and determine all retained single-trace data from five normalized double-trace coefficients, reproducing characteristic zeros of the heterotic expansion. The gravity-decoupled branch with $a_{0,0}\neq0$ instead recovers the open-superstring constraints through the orders studied. For $SO(N)$, dispersive positivity restricts the gravitational branch to a narrow band closely bounded by rescalings of the heterotic and simplest IST amplitudes, together with a broader region near the origin. Within the numerical truncations, this region contracts as $N$ increases; at $N=49$, the extended band disappears, although a small region near the origin survives.

An analytic derivation of the heterotic four-point amplitude under suitable ultraviolet assumptions could follow the strategies of~\cite{Sung:2026euf,Huang:2026ghg}. Both use special six-point kinematics to eliminate unknown regular contributions and turn supersymmetry, scalar parity, and factorization into all-order functional relations. Sung combines a reconstruction equation with a low-order SUSY relation and spectral positivity to establish Veneziano uniqueness; ref.~\cite{Huang:2026ghg} derives gauge and gravity master equations using scalar parity and a discrete R-symmetry, determines their exponential solutions, and reconstructs the amplitudes from regular forward data. Extending these methods to the mixed- and triple-trace sectors could establish the common exponential form factor and rational double-trace prefactor to all orders.

The appearance of IST solutions here and in the maximal-supergravity bootstrap may reflect the common odd-zeta structure of closed-string four-point form factors. The nonlinear constraints leave the coefficients occupying the odd-zeta positions as free parameters, admitting solutions other than their string values. Whether the harmonic truncation defining the four-point IST solution extends to higher multiplicity, where multiple zeta values occur, remains to be determined. Compatibility with their algebraic relations and with factorization would provide a route to consistent $n$-point IST amplitudes.

\acknowledgments
We thank Justin Berman, Aditi V. Chandra, Shi Chen, Nicholas Geiser, Loki L. Lin, Roger Morales, Shruti Paranjape, David Poland, Francesco Riva, and Marcos Skowronek for valuable discussions.
The research of Y-tH is supported by Taiwan NSTC grant 115-2112-M-002 -035 -MY3. The research of H-CW is supported by the US Department of Energy under contract DE-SC0010010 Task F and by Simons Investigator Award \#376208. L-YC is supported by DOE grant DE-SC0017660. Computations were performed using the Yale Grace computing cluster, supported by the facilities and staff of the Yale University Faculty of Sciences High Performance Computing Center, and the resources and services of the Center for Computation and Visualization at Brown University. Large language models were used to assist with code implementation, data visualization, and manuscript preparation, all of which were reviewed and verified by the authors.

\appendix
\section{Single-trace limit and equivalence to color-ordered positivity}\label{sec:single_trace_limit}

For a purely single-trace amplitude, positivity is usually imposed on the Gegenbauer coefficients of a color-ordered partial amplitude. We show that this is equivalent to positivity in the color-projector basis.

Start from a single-trace dispersive representation
\begin{equation}
A^{\rm ST}(s,t)=\sum_{J}\int_{\Lambda^2}^{\infty}ds'\,\frac{\tilde\rho_{J}(s')\,G_{J}(1+2t/s')}{s'-s}\, .
\end{equation}
Using the trace decomposition \eqref{eq:single_double_tr}, the $s$-channel discontinuity of the full single-trace amplitude is
\begin{equation}
\operatorname{Disc}_s A_4
=\operatorname{tr}(1234)\operatorname{Disc}_s A^{\rm ST}(s,t)
+\operatorname{tr}(1342)\operatorname{Disc}_s A^{\rm ST}(u,s)\, .
\end{equation}
The third single-trace ordering contributes to the crossed cut.  Since
\begin{equation}
G_J\left(1+\frac{2u}{s}\right)=G_J\left(-1-\frac{2t}{s}\right)=(-1)^J G_J\left(1+\frac{2t}{s}\right)\,,
\end{equation}
one obtains
\begin{equation}\label{eq:single_trace_in_projector_basis}
\operatorname{Disc}_s A_4
=\sum_J\tilde\rho_J(s)G_J(\cos\theta)\,
\vec{\mathbb T}\cdot
\begin{pmatrix}
1\\ (-1)^J\\ 0\\ 0\\ 0\\ 0
\end{pmatrix}
=\sum_J\tilde\rho_{J}(s)\,G_{J}(\cos\theta)\,\vec{\mathbb P}^{\,s}\cdot
\begin{pmatrix}
\frac{N-1}{2}\bigl(1+(-1)^J\bigr) \\
\frac{N-2}{4}\bigl(1+(-1)^J\bigr) \\
0 \\
0 \\
\frac{N-2}{4}\bigl(1-(-1)^J\bigr) \\
0
\end{pmatrix} .
\end{equation}
Thus $\tilde\rho_J(s)\geq0$ implies projector positivity. Even spins populate the singlet and symmetric traceless representations with spectral densities in the ratio $2(N-1)/(N-2)$; odd spins populate only the adjoint.

Conversely, take nonnegative projector spectral densities
\begin{equation}
\vec\rho_J=(\rho_1,\rho_2,\rho_3,\rho_4,\rho_5,\rho_6)^{\rm T}\,,\qquad \rho_R\geq 0\, ,
\end{equation}
and require the double-trace discontinuity to vanish in both the $s$ and $u$ channels. The last three components in the trace basis must then vanish. The basis transformations give
\begin{align}
M_s^T
\begin{pmatrix}
\rho_1\\ \rho_2\\ \rho_3\\ \rho_4\\ \rho_5\\ \rho_6
\end{pmatrix}
&=
\begin{pmatrix}
\dfrac{2(\rho_2-\rho_3+\rho_5-\rho_6)}{N-2}\\[1.2ex]
\dfrac{2(\rho_2-\rho_3-\rho_5+\rho_6)}{N-2}\\[1.2ex]
\dfrac{2(\rho_3-\rho_4)}{3}\\[1.2ex]
\dfrac{2\big((N-2)\rho_1-2(N-1)\rho_2+N\rho_3\big)}{(N-2)(N-1)N}\\[1.2ex]
\dfrac{2\rho_3+\rho_4-3\rho_6}{6}\\[1.2ex]
\dfrac{2\rho_3+\rho_4+3\rho_6}{6}
\end{pmatrix}, \\
M_s^T M_{us}^T
\begin{pmatrix}
\rho_1\\ \rho_2\\ \rho_3\\ \rho_4\\ \rho_5\\ \rho_6
\end{pmatrix}
&=
\begin{pmatrix}
\dfrac{2(\rho_3-\rho_4)}{3}\\[1.2ex]
\dfrac{2(\rho_2-\rho_3-\rho_5+\rho_6)}{N-2}\\[1.2ex]
\dfrac{2(\rho_2-\rho_3+\rho_5-\rho_6)}{N-2}\\[1.2ex]
\dfrac{2\rho_3+\rho_4-3\rho_6}{6}\\[1.2ex]
\dfrac{2\big((N-2)\rho_1-2(N-1)\rho_2+N\rho_3\big)}{(N-2)(N-1)N}\\[1.2ex]
\dfrac{2\rho_3+\rho_4+3\rho_6}{6}
\end{pmatrix} .
\end{align}
Here $M_{us} = M_{su}^{-1}$ maps the s-channel projector to the u-channel projectors. The double-trace entries vanish when
\begin{equation}\label{eq:no_double_tr_constraint}
0=\rho_6=(N-2)\rho_1-2(N-1)\rho_2+N\rho_3=2\rho_3+\rho_4\, .
\end{equation}
Nonnegativity of the $\rho_R$ then forces $\rho_3=\rho_4=\rho_6=0$, leaving
\begin{equation}\label{eq:no_double_tr_spec_density_pre_parity}
\vec\rho_J=\left(\frac{N-1}{2}\alpha,\,\frac{N-2}{4}\alpha,\,0,\,0,\,\beta,\,0\right),\qquad \alpha,\beta\geq 0\, .
\end{equation}
Bose symmetry assigns the symmetric representations $R=1,2$ to even spin and the antisymmetric representation $R=5$ to odd spin, giving
\begin{equation}\label{eq:no_double_tr_spec_density}
\vec\rho_J=
\left(
\frac{N-1}{2}\frac{1+(-1)^J}{2}\alpha,
\frac{N-2}{4}\frac{1+(-1)^J}{2}\alpha,
0,
0,
\frac{1-(-1)^J}{2}\beta,
0
\right),\qquad \alpha,\beta\geq 0\, .
\end{equation}
This agrees with~\eqref{eq:single_trace_in_projector_basis} up to positive normalizations of the even- and odd-spin densities. Recovering the single-trace EFT-hedron therefore requires cyclicity and the full vanishing of the double-trace discontinuity.

\section{Unitarity of the $q=1$ IST amplitude}
\label{sec:Unitarity_IST}

We derive the unitarity bounds on the $q=1$ IST amplitude for integer $N\geq4$ by expanding its massive $s$-channel residue in the six $SO(N)$ projectors. We use $m=M_{\rm gap}$ and the physical coupling ratio $R_{\rm gap}=g^2/(\kappa^2m^2)=xr$. The amplitude is
\begin{equation}
\begin{aligned}
    F^{\text{ST}}(s,t) &= -\frac{g^2}{s t}\frac{(m^2+s)(m^2+t)(m^2+u)}{(m^2-s)(m^2-t)(m^2-u)},\\
    F^{\text{DT}}(s,t) &= -\frac{m^2\kappa^2}{s(m^2+s)}\frac{(m^2+s)(m^2+t)(m^2+u)}{(m^2-s)(m^2-t)(m^2-u)}.
\end{aligned}
\end{equation}
The massive $s$-channel pole is at $s=m^2$. Using $M_s$ in~\eqref{eq:Ms_trace_matrix}, we express its residue in the projector basis as
\begin{equation}
\begin{aligned}\label{eq:IST_residue}
    \vec{\rho}=&\kappa^2\left(-\frac{4 \left(N(N-1)+(N-2)R_{\rm gap}+4\right)}{(c-3) (c+3) }-\frac{N(N-1)}{2},\right.\\
    &\qquad -\frac{2 \left((N-4)R_{\rm gap}+8\right)}{(c-3) (c+3)},\frac{4 (R_{\rm gap}-4)}{(c-3) (c+3)},\\
    &\qquad\left.-\frac{8 (R_{\rm gap}+2)}{(c-3) (c+3)},-\frac{2 c (N-2)R_{\rm gap}}{(c-3) (c+3)},0\right)^T,
\end{aligned}
\end{equation}
where $c=\cos\theta=1+2t/m^2$. We obtain the partial-wave expansion from Heine's formula~\cite{NIST:Heine},
\begin{equation}
    \frac{1}{y-x} =\sum_{J\geq 0} (2J+1)Q_J(y)P_J(x),
\end{equation}
where $P_J$ and $Q_J$ are the Legendre polynomial and function of the second kind. Partial fractions give
\begin{equation}
\begin{aligned}\label{eq:Heine}
    &\frac{1}{(c-3)(c+3)} = -\frac{1}{3}\sum_{\substack{J\geq 0,\\J\text{ even}}}(2J+1)Q_J(3)P_J(c),\\
    &\frac{c}{(c-3)(c+3)} = -\sum_{\substack{J\geq 0,\\J\text{ odd}}}(2J+1)Q_J(3)P_J(c).
\end{aligned}
\end{equation}
The coefficients $Q_J(3)$ are positive for every $J\geq0$.

Combining~\eqref{eq:IST_residue} and~\eqref{eq:Heine} gives
\begin{equation}
\begin{aligned}
    &\rho^1 = \kappa^2\sum_{\substack{J\geq 0,\\J\text{ even}}}\left[(2J+1)\frac{4 \left(N(N-1)+(N-2)R_{\rm gap}+4\right)}{3}Q_J(3)\right.\\
    &\hspace{42mm}\left.-\delta_{J,0}\frac{N(N-1)}{2}\right]P_J(c),\\
    &\rho^2= \kappa^2\sum_{\substack{J\geq 0,\\J\text{ even}}}(2J+1)\frac{2 \left((N-4)R_{\rm gap}+8\right)}{3}Q_J(3)P_J(c),\\
    &\rho^3= \kappa^2\sum_{\substack{J\geq 0,\\J\text{ even}}}(2J+1)\frac{4 (4-R_{\rm gap})}{3}Q_J(3)P_J(c),\\
    &\rho^4= \kappa^2\sum_{\substack{J\geq 0,\\J\text{ even}}}(2J+1)\frac{8 (R_{\rm gap}+2)}{3}Q_J(3)P_J(c),\\
    &\rho^5= \kappa^2\sum_{\substack{J\geq 0,\\J\text{ odd}}}2(2J+1) (N{-}2)R_{\rm gap}\;Q_J(3)P_J(c),\\
    &\rho^6= 0.
\end{aligned}
\end{equation}
The first four representations contain only even spins, and the fifth only odd spins, as required by Bose symmetry. Together with $g^2\geq0$, positivity in the third representation restricts $R_{\rm gap}$ to $0\leq R_{\rm gap}\leq4$. For $N\geq4$, every coefficient in this range is nonnegative except possibly the scalar in the singlet channel. Requiring $\rho^1_{J=0}\geq0$ and taking $g\neq0$ gives
\begin{equation}
\begin{aligned}
    &4\leq N \leq 7:\quad &0< R_{\rm gap}\leq 4,\\
    &8\leq N \leq 48: &\frac{3 N(N{-}1){-}(4 N(N{-}1)+16)\log (2)}{(4 N{-}8) \log (2)} \leq R_{\rm gap}\leq 4.
\end{aligned}
\end{equation}
The $q=1$ IST amplitude is non-unitary for $N\geq49$. The endpoint $R_{\rm gap}=0$ also satisfies these partial-wave positivity conditions for $4\leq N\leq7$; it is omitted above because we assume a nonzero gauge coupling.

\bibliographystyle{JHEP}
\bibliography{biblio}

@article{Tolley:2020gtv,
    author = "Tolley, Andrew J. and Wang, Zi-Yue and Zhou, Shuang-Yong",
    title = "{New positivity bounds from full crossing symmetry}",
    eprint = "2011.02400",
    archivePrefix = "arXiv",
    primaryClass = "hep-th",
    doi = "10.1007/JHEP05(2021)255",
    journal = "JHEP",
    volume = "05",
    pages = "255",
    year = "2021"
}

@article{Bachu:2022gof,
    author = "Bachu, Brad and Hillman, Aaron",
    title = "{Stringy Completions of the Standard Model from the Bottom Up}",
    eprint = "2212.03871",
    archivePrefix = "arXiv",
    primaryClass = "hep-th",
    month = "12",
    year = "2022"
}

@article{Caron-Huot:2021rmr,
    author = "Caron-Huot, Simon and Mazac, Dalimil and Rastelli, Leonardo and Simmons-Duffin, David",
    title = "{Sharp boundaries for the swampland}",
    eprint = "2102.08951",
    archivePrefix = "arXiv",
    primaryClass = "hep-th",
    doi = "10.1007/JHEP07(2021)110",
    journal = "JHEP",
    volume = "07",
    pages = "110",
    year = "2021"
}

@article{Sinha:2020win,
    author = "Sinha, Aninda and Zahed, Ahmadullah",
    title = "{Crossing Symmetric Dispersion Relations in Quantum Field Theories}",
    eprint = "2012.04877",
    archivePrefix = "arXiv",
    primaryClass = "hep-th",
    doi = "10.1103/PhysRevLett.126.181601",
    journal = "Phys. Rev. Lett.",
    volume = "126",
    number = "18",
    pages = "181601",
    year = "2021"
}

@article{Chiang:2022jep,
    author = "Chiang, Li-Yuan and Huang, Yu-tin and Li, Wei and Rodina, Laurentiu and Weng, He-Chen",
    title = "{(Non)-projective bounds on gravitational EFT}",
    eprint = "2201.07177",
    archivePrefix = "arXiv",
    primaryClass = "hep-th",
    month = "1",
    year = "2022"
}

@article{deRham:2017avq,
    author = "de Rham, Claudia and Melville, Scott and Tolley, Andrew J. and Zhou, Shuang-Yong",
    title = "{Positivity bounds for scalar field theories}",
    eprint = "1702.06134",
    archivePrefix = "arXiv",
    primaryClass = "hep-th",
    doi = "10.1103/PhysRevD.96.081702",
    journal = "Phys. Rev. D",
    volume = "96",
    number = "8",
    pages = "081702",
    year = "2017"
}

@article{Bellazzini:2020cot,
    author = "Bellazzini, Brando and Elias Mir{\'o}, Joan and Rattazzi, Riccardo and Riembau, Marc and Riva, Francesco",
    title = "{Positive moments for scattering amplitudes}",
    eprint = "2011.00037",
    archivePrefix = "arXiv",
    primaryClass = "hep-th",
    doi = "10.1103/PhysRevD.104.036006",
    journal = "Phys. Rev. D",
    volume = "104",
    number = "3",
    pages = "036006",
    year = "2021"
}

@article{Huang:2020nqy,
    author = "Huang, Yu-tin and Liu, Jin-Yu and Rodina, Laurentiu and Wang, Yihong",
    title = "{Carving out the Space of Open-String S-matrix}",
    eprint = "2008.02293",
    archivePrefix = "arXiv",
    primaryClass = "hep-th",
    reportNumber = "NCTS-TH/2011",
    doi = "10.1007/JHEP04(2021)195",
    journal = "JHEP",
    volume = "04",
    pages = "195",
    year = "2021"
}

@article{Sung:2026euf,
    author = "Sung, Ilmo",
    title = "{Six-point consistency and uniqueness of the Veneziano amplitude}",
    eprint = "2609.29919",
    archivePrefix = "arXiv",
    primaryClass = "hep-th",
    month = "9",
    year = "2026"
}

@article{Huang:2026ghg,
    author = "Huang, Yu-tin and Wan, Shi-Lin and Wang, Zhuo-Hui and Zhou, Shuang-Yong",
    title = "{Infrared Consistency and the Uniqueness of String Amplitudes}",
    eprint = "2610.02124",
    archivePrefix = "arXiv",
    primaryClass = "hep-th",
    month = "10",
    year = "2026"
}

@article{Cheung:2025tbr,
    author = "Cheung, Clifford and Remmen, Grant N. and Sciotti, Francesco and Tarquini, Michele",
    title = "{Strings from Almost Nothing}",
    eprint = "2508.09246",
    archivePrefix = "arXiv",
    primaryClass = "hep-th",
    reportNumber = "CALT-TH 2025-027",
    doi = "10.1103/cw4p-cqh7",
    journal = "Phys. Rev. Lett.",
    volume = "136",
    number = "25",
    pages = "251601",
    year = "2026"
}

@article{Berman:2025owb,
    author = "Berman, Justin and Elvang, Henriette and Figueiredo, Carolina",
    title = "{Splitting regions and shrinking islands from higher point constraints}",
    eprint = "2506.22538",
    archivePrefix = "arXiv",
    primaryClass = "hep-th",
    reportNumber = "LITP-25-08",
    doi = "10.1007/JHEP10(2025)226",
    journal = "JHEP",
    volume = "10",
    pages = "226",
    year = "2025"
}

@article{Chiang:2023quf,
    author = "Chiang, Li-Yuan and Huang, Yu-tin and Weng, He-Chen",
    title = "{Bootstrapping string theory EFT}",
    eprint = "2310.10710",
    archivePrefix = "arXiv",
    primaryClass = "hep-th",
    doi = "10.1007/JHEP05(2024)289",
    journal = "JHEP",
    volume = "05",
    pages = "289",
    year = "2024"
}

@article{Berman:2023jys,
    author = "Berman, Justin and Elvang, Henriette and Herderschee, Aidan",
    title = "{Flattening of the EFT-hedron: supersymmetric positivity bounds and the search for string theory}",
    eprint = "2310.10729",
    archivePrefix = "arXiv",
    primaryClass = "hep-th",
    doi = "10.1007/JHEP03(2024)021",
    journal = "JHEP",
    volume = "03",
    pages = "021",
    year = "2024"
}

@article{Elvang:2026pmc,
    author = "Elvang, Henriette and Herderschee, Aidan and Morales, Roger",
    title = "{String theory from maximal supersymmetry}",
    eprint = "2601.11705",
    archivePrefix = "arXiv",
    primaryClass = "hep-th",
    reportNumber = "LITP-25-18",
    doi = "10.1007/JHEP07(2026)105",
    journal = "JHEP",
    volume = "07",
    pages = "105",
    year = "2026"
}

@article{Berman:2026ezk,
    author = "Berman, Justin and Caron-Huot, Simon and Chandra, Aditi V. and Elvang, Henriette and Herderschee, Aidan and Lin, Loki L. and Morales, Roger",
    title = "{Gravitational Effective Theories with Maximal Supersymmetry and a Peculiar Parity}",
    eprint = "2607.14230",
    archivePrefix = "arXiv",
    primaryClass = "hep-th",
    reportNumber = "LITP-26-13",
    month = "7",
    year = "2026"
}

@article{Elvang:2009wd,
    author = "Elvang, Henriette and Freedman, Daniel Z. and Kiermaier, Michael",
    title = "{Solution to the Ward Identities for Superamplitudes}",
    eprint = "0911.3169",
    archivePrefix = "arXiv",
    primaryClass = "hep-th",
    reportNumber = "MIT-CTP-4086, PUPT-2323, MCTP-09-53",
    doi = "10.1007/JHEP10(2010)103",
    journal = "JHEP",
    volume = "10",
    pages = "103",
    year = "2010"
}

@article{Elvang:2010xn,
    author = "Elvang, Henriette and Freedman, Daniel Z. and Kiermaier, Michael",
    title = "{SUSY Ward identities, Superamplitudes, and Counterterms}",
    eprint = "1012.3401",
    archivePrefix = "arXiv",
    primaryClass = "hep-th",
    reportNumber = "MCTP-10-56, MIT-CTP-4197, PUPT-2361",
    doi = "10.1088/1751-8113/44/45/454009",
    journal = "J. Phys. A",
    volume = "44",
    pages = "454009",
    year = "2011"
}

@misc{SparseRREF,
  author    = {Li, Zhenjie},
  title     = {{SparseRREF}: Exact Sparse Reduced Row Echelon Form in {C++}},
  version   = {0.4.0},
  note      = {Version 0.4.0},
  doi       = {10.5281/zenodo.21278863},
  url       = {https://github.com/munuxi/SparseRREF}
}

@misc{Headrick:grassmann,
  author       = {Matthew Headrick},
  title        = {{grassmann.m}: A Mathematica package for algebra and calculus with Grassmann variables},
  howpublished = {\url{https://sites.google.com/view/matthew-headrick/mathematica}}
}

@article{Schlotterer:2012ny,
    author = "Schlotterer, O. and Stieberger, S.",
    title = "{Motivic Multiple Zeta Values and Superstring Amplitudes}",
    eprint = "1205.1516",
    archivePrefix = "arXiv",
    primaryClass = "hep-th",
    reportNumber = "AEI-2012-039, MPP-2012-859",
    doi = "10.1088/1751-8113/46/47/475401",
    journal = "J. Phys. A",
    volume = "46",
    pages = "475401",
    year = "2013"
}

@article{BoisvertKnopRastelli,
    author = "Boisvert, Mathieu and Knop, Waltraut and Rastelli, Leonardo",
    title = "{Where is tree-level heterotic string theory?}",
    eprint = "2606.09980",
    archivePrefix = "arXiv",
    primaryClass = "hep-th",
    reportNumber = "DESY-26-072",
    month = "6",
    year = "2026"
}

@article{Caron-Huot:2020cmc,
    author = "Caron-Huot, Simon and Van Duong, Vincent",
    title = "{Extremal Effective Field Theories}",
    eprint = "2011.02957",
    archivePrefix = "arXiv",
    primaryClass = "hep-th",
    doi = "10.1007/JHEP05(2021)280",
    journal = "JHEP",
    volume = "05",
    pages = "280",
    year = "2021"
}

@article{Simmons-Duffin:2015qma,
    author = "Simmons-Duffin, David",
    title = "{A Semidefinite Program Solver for the Conformal Bootstrap}",
    eprint = "1502.02033",
    archivePrefix = "arXiv",
    primaryClass = "hep-th",
    doi = "10.1007/JHEP06(2015)174",
    journal = "JHEP",
    volume = "06",
    pages = "174",
    year = "2015"
}

@article{Landry:2019qug,
    author = "Landry, Walter and Simmons-Duffin, David",
    title = "{Scaling the semidefinite program solver SDPB}",
    eprint = "1909.09745",
    archivePrefix = "arXiv",
    primaryClass = "hep-th",
    reportNumber = "CALT-TH 2019-038",
    month = "9",
    year = "2019"
}

@article{Albert:2024yap,
    author = "Albert, Jan and Knop, Waltraut and Rastelli, Leonardo",
    title = "{Where is tree-level string theory?}",
    eprint = "2406.12959",
    archivePrefix = "arXiv",
    primaryClass = "hep-th",
    reportNumber = "YITP-SB-2024-12",
    doi = "10.1007/JHEP02(2025)157",
    journal = "JHEP",
    volume = "02",
    pages = "157",
    year = "2025"
}

@article{Albert:2022oes,
    author = "Albert, Jan and Rastelli, Leonardo",
    title = "{Bootstrapping pions at large N}",
    eprint = "2203.11950",
    archivePrefix = "arXiv",
    primaryClass = "hep-th",
    reportNumber = "YITP-SB-2022-07",
    doi = "10.1007/JHEP08(2022)151",
    journal = "JHEP",
    volume = "08",
    pages = "151",
    year = "2022"
}

@article{Albert:2023seb,
    author = "Albert, Jan and Henriksson, Johan and Rastelli, Leonardo and Vichi, Alessandro",
    title = "{Bootstrapping mesons at large N: Regge trajectory from spin-two maximization}",
    eprint = "2312.15013",
    archivePrefix = "arXiv",
    primaryClass = "hep-th",
    reportNumber = "YITP-SB-2023-41",
    doi = "10.1007/JHEP09(2024)172",
    journal = "JHEP",
    volume = "09",
    pages = "172",
    year = "2024"
}

@misc{NIST:Heine,
  title        = "{NIST Digital Library of Mathematical Functions}, {\S}14.28(ii)",
  howpublished = "\url{https://dlmf.nist.gov/14.28.E2}",
  key          = "DLMF"
}

@article{ArkaniHamedHuangHuang2021EFT,
    author = "Arkani-Hamed, Nima and Huang, Tzu-Chen and Huang, Yu-tin",
    title = "{The EFT-Hedron}",
    eprint = "2012.15849",
    archivePrefix = "arXiv",
    primaryClass = "hep-th",
    reportNumber = "NCTS-TH/2014, CALT-TH 2020-061",
    doi = "10.1007/JHEP05(2021)259",
    journal = "JHEP",
    volume = "05",
    pages = "259",
    year = "2021"
}

@article{BernKosmopoulosZhiboedov2021,
    author = "Bern, Zvi and Kosmopoulos, Dimitrios and Zhiboedov, Alexander",
    title = "{Gravitational Effective Field Theory Islands, Low-Spin Dominance, and the Four-Graviton Amplitude}",
    eprint = "2103.12728",
    archivePrefix = "arXiv",
    primaryClass = "hep-th",
    reportNumber = "CERN-TH-2021-035",
    doi = "10.1088/1751-8121/ac0e51",
    journal = "J. Phys. A",
    volume = "54",
    number = "34",
    pages = "344002",
    year = "2021"
}

@article{CalistoEtAl2025Completeness,
    author = "Calisto, Francesco and Cheung, Clifford and Remmen, Grant N. and Sciotti, Francesco and Tarquini, Michele",
    title = "{Completeness from Gravitational Scattering}",
    eprint = "2512.11955",
    archivePrefix = "arXiv",
    primaryClass = "hep-th",
    reportNumber = "CALT-TH 2025-040",
    doi = "10.1103/pztp-rchl",
    journal = "Phys. Rev. D",
    volume = "113",
    pages = "106008",
    year = "2026"
}

@article{HillmanHuangRodinaRumbutis2024,
    author = "Hillman, Aaron and Huang, Yu-tin and Rodina, Laurentiu and Rumbutis, Justinas",
    title = "{Spectral Constraints on Theories of Colored Particles and Gravity}",
    eprint = "2411.04857",
    archivePrefix = "arXiv",
    primaryClass = "hep-th",
    doi = "10.1103/k23f-y47w",
    journal = "Phys. Rev. Lett.",
    volume = "135",
    number = "6",
    pages = "061604",
    year = "2025"
}

@article{Wan:2026pjq,
    author = "Wan, Shi-Lin and Zhou, Shuang-Yong",
    title = "{Analytic Bootstrap of the Veneziano Amplitude}",
    eprint = "2605.11084",
    archivePrefix = "arXiv",
    primaryClass = "hep-th",
    reportNumber = "USTC-ICTS/PCFT-26-28",
    month = "5",
    year = "2026"
}

@article{Berman:2024kdh,
    author = "Berman, Justin",
    title = "{Analytic bounds on the spectrum of crossing symmetric S-matrices}",
    eprint = "2410.01914",
    archivePrefix = "arXiv",
    primaryClass = "hep-th",
    reportNumber = "LCTP-24-16",
    doi = "10.1007/JHEP08(2025)066",
    journal = "JHEP",
    volume = "08",
    pages = "066",
    year = "2025"
}
\end{document}